\documentclass[pre,reprint,secnumarabic,nofootinbib,floatfix,superscriptaddress,
amsmath,amssymb,showkeys,a4paper] {revtex4-1}

\usepackage{graphicx}
\usepackage[usenames, dvipsnames]{color}
\usepackage{epsfig}
\usepackage{wasysym}

\begin{document}

\title{Topographical pathways guide chemical microswimmers}

\author{Juliane Simmchen}

\affiliation{Max-Planck-Institut f\"ur Intelligente Systeme, Heisenbergstr.~3, 
70569 Stuttgart, Germany}

\author{Jaideep Katuri}

\affiliation{Max-Planck-Institut f\"ur Intelligente Systeme, Heisenbergstr.~3, 
70569 Stuttgart, Germany}

\affiliation{Institut de Bioenginyeria de Catalunya (IBEC), Baldiri I Reixac 10-12, 
08028 Barcelona, Spain}

\author{William E. Uspal}
 
\author{Mihail N. Popescu}

\author{Mykola Tasinkevych}

\email{\texttt{miko@is.mpg.de}}

\affiliation{Max-Planck-Institut f\"ur Intelligente Systeme, Heisenbergstr.~3, 
70569 Stuttgart, Germany}

\affiliation{IV. Institut f\"ur Theoretische Physik, Universit\"{a}t Stuttgart,
Pfaffenwaldring 57, D-70569 Stuttgart, Germany}

\author{Samuel S{\'a}nchez}

\email{\texttt{ssanchez@ibecbarcelona.eu}}

\affiliation{Max-Planck-Institut f\"ur Intelligente Systeme, Heisenbergstr.~3, 
70569 Stuttgart, Germany}

\affiliation{Institut de Bioenginyeria de Catalunya (IBEC), Baldiri I Reixac 10-12, 
08028 Barcelona, Spain}

\affiliation{Instituci{\'o} Catalana de Recerca i Estudis Avancats (ICREA), 
Pg. Llu{\'i}s Companys 23, 08010,Barcelona, Spain}

\date{\today}

\begin{abstract}
Achieving control over the directionality of active colloids is essential for their 
use in practical applications such as cargo carriers in microfluidic devices. So far, 
guidance of spherical Janus colloids was mainly realized using specially engineered 
magnetic multilayer coatings combined with external magnetic fields. Here, we 
demonstrate that step-like sub-micron topographical features can be used as reliable 
docking and guiding devices for chemically active spherical Janus colloids. For various 
topographic features (stripes, squares or circular posts) docking of the colloid at the 
feature edge is robust and reliable. Furthermore, the colloids move along the edges for 
significantly long times, which systematically increase with fuel concentration. The 
observed phenomenology is qualitatively captured by a simple continuum model of 
self-diffusiophoresis near confining boundaries, indicating that the chemical activity 
and associated hydrodynamic interactions with the nearby topography are the main 
physical ingredients behind the observed behaviour.
\end{abstract}



\maketitle

\section{\label{intro} Introduction}

Catalytically active micron-sized objects can self-propel by various mechanisms, 
including bubble ejection, diffusio-, and electro-phoresis, when parts of their surface 
catalyze a chemical reaction in a surrounding liquid. In future, such chemically active 
micromotors may serve as autonomous carriers working within microfluidic devices to 
fulfill complex tasks \cite{ref1,ref2,ref3}. However, in order to achieve this goal, it 
is essential to gain robust control over the directionality of particle motion. Although 
it has been more than a decade since motile chemically active colloids were first 
reported \cite{ref4,ref5,ref6,ref7} this remains a challenging issue, in particular for 
the case of spherical particles.

Two main methods of guidance have been so far employed with varying degrees of success. 
The first one uses controlled spatial gradients of “fuel” concentration. This approach 
suffers, however, from severe difficulties in creating and maintaining chemical 
gradients, and the spatial precision of guidance remains rather poor 
\cite{ref8,ref9,ref10,ref11,ref12}. The second approach relies upon the use of external 
magnetic fields in combination with particles with suitably designed magnetic coatings or 
inclusions \cite{ref5,ref13}. This proved to be a very precise guidance mechanism which 
could be employed straightforwardly for the case of rod-like particles \cite{ref14} but 
difficult to extend to the case of spherical colloids, where it requires sophisticated
engineering of multilayer magnetic coatings \cite{ref7,ref15,ref16,ref17}. Additionally, 
individualized guidance of specific particles is difficult to achieve without complicated 
external apparatus and feedback loops \cite{ref18}. The advantages of autonomous 
operation are thereby significantly hindered.

While these methods are quite general in their applicability, we note here that the 
synthetic micromotors are, in general, density mismatched with the suspending medium and 
therefore tend to sediment and move near surfaces. Furthermore, even in situations in 
which sedimentation can be neglected (e.g., in the case of neutrally buoyant swimmers) 
the presence of confining surfaces has profound consequences on swimming trajectories, as 
discussed below. Theoretical studies have shown that long range hydrodynamic interactions 
between microswimmers and nearby surfaces \cite{ref19,ref20} can give rise to trapping at 
the walls or circular motion. Moreover, a theoretical study of a model active Janus
colloid moving near a planar inert wall has revealed complex behaviour, including novel 
sliding and hovering steady states \cite{ref21}. Experimentally, wall-bounded motion of 
active Janus particles was evidenced in the study by Bechinger \textit{et al.} 
\cite{ref22}, while capture into orbital trajectories of active bi-metallic rods by large
spherical beads or of Janus colloids in colloidal crystals has recently been reported 
\cite{ref23,takagi2014}. Capture of microswimmers by spherical obstacles via hydrodynamic 
interactions has been modeled theoretically by Lauga \textit{et al.} \cite{ref24}.

\begin{figure*}[!thb]
\includegraphics[width = 0.9\textwidth]{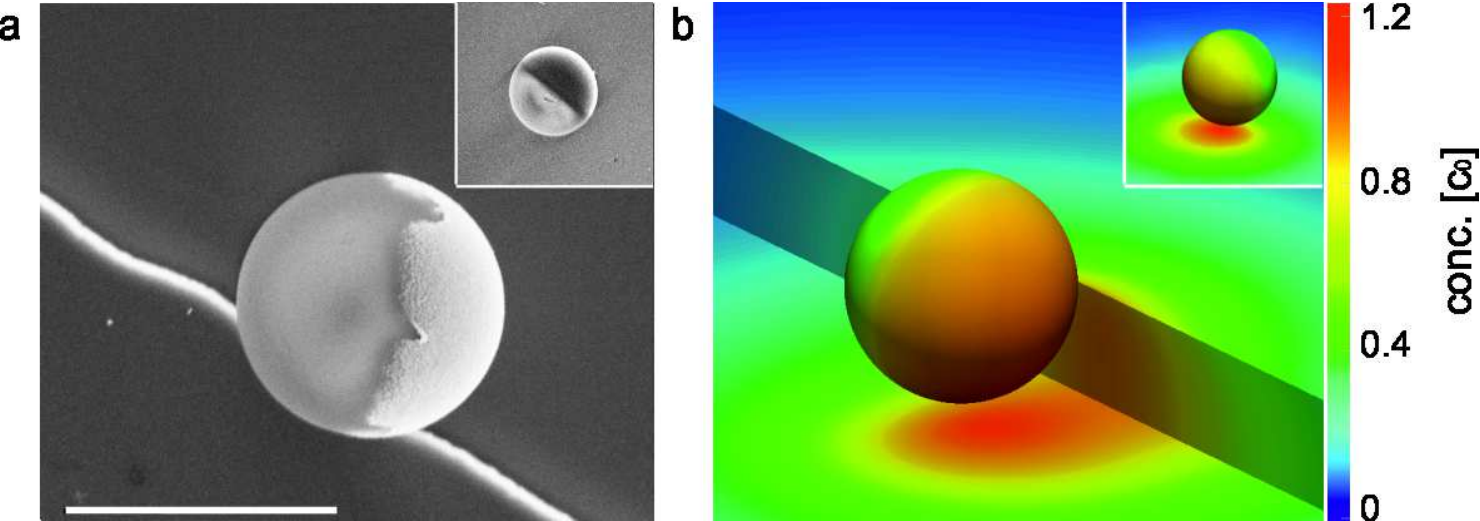}
\caption{\label{fig1} Janus microswimmers near sub-micron steps. (a) Top view SEM image 
of a spherical Janus motor on a Silicon substrate with a Silicon step. The lighter part 
of the Janus particle corresponds to Pt, while the gray part is the SiO$_2$; scale bar 
corresponds to 2 $\mathrm{\mu m}$. (b) Color-coded steady-state distribution 
$c(\mathbf{r})$ of reaction products around a half covered Janus particle at an inert 
wall and a step with height $h_{step} = R$, where $R$ is the particle radius. The color 
map shows $c(\mathbf{r})$ at the surfaces of the particle and substrate, and is 
represented in units of $c_0$ defined in Methods. The insets show the particles on a flat
surface.}
\end{figure*}
This intrinsic tendency of the active swimmers to operate near bounding surfaces 
motivated us to examine whether it can be further exploited to achieve directional 
guidance of chemically active microswimmers by endowing the wall with small height 
step-like topographical features, as shown in Fig. \ref{fig1}, which the particles can 
eventually exploit as pathways. Recently, Palacci \textit{et al.} have shown that shallow 
rectangular grooves can efficiently guide photocatalytic hematite swimmers that have size 
comparable with the width of the groove \cite{ref25}. Because of the strong lateral 
confinement, it is hard to
discriminate between the different physical contributions which lead to particle 
guidance. Here we use a much less restrictive geometry -- a shallow topographical step -- 
and it is a priori not clear whether a self-phoretic swimmer can follow such features. We 
report experimental evidence that Janus microswimmers can follow step-like topographical 
features that are only a fraction of the particle radius in height. This is, in some 
sense, similar to the strategy employed in natural systems well below the microscale: 
within cells, protein motors such as myosin, kinesin and dynein use binding to 
microtubules to switch to directional motion \cite{ref26,ref27}. The guidance of 
microswimmers through patterned device topography that we propose and demonstrate in this 
study may pave the way for new methods of self-propeller motion control based upon 
patterned walls.

\section{\label{results} Results}

\subsection{Dynamics of Janus microswimmers at a planar wall}

Janus particles are fabricated by vapor deposition of a thin layer of Pt (7 nm) on 
SiO$_2$ particles (diameters of approximately 2 and 5 $\mathrm{\mu m}$). For details on 
the fabrication see Methods section. Scanning electron microscopy (SEM) images of a 
Janus particle at the step edge are shown in Fig. \ref{fig1}a, while Fig. \ref{fig1}b 
illustrates numerically calculated steady-state distributions of the reaction products 
around a model half Pt-covered Janus sphere near a step.
\begin{figure*}[!thb]
\includegraphics[width = 0.9\textwidth]{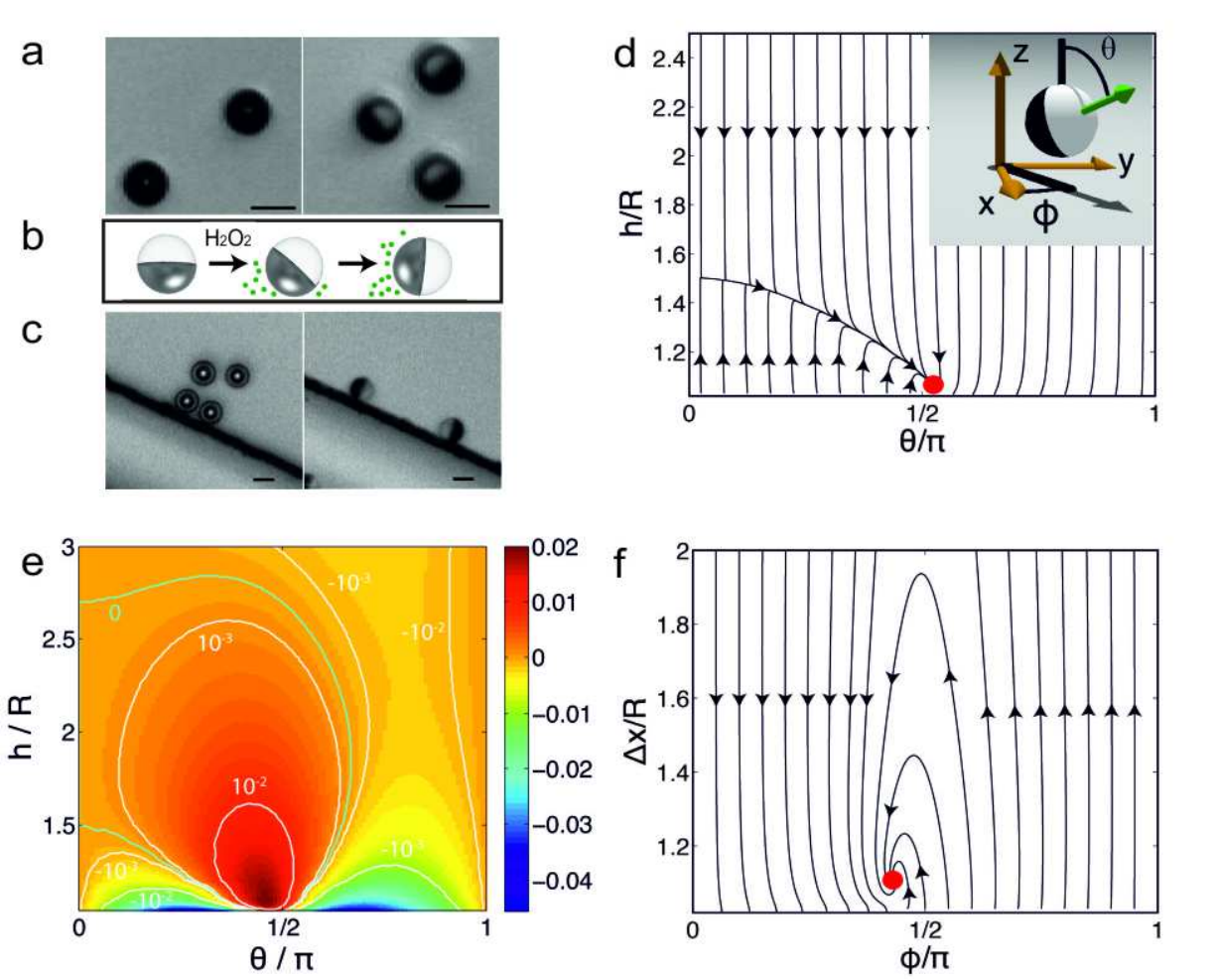}
\caption{\label{fig2} Behavior of Janus particles near planar surfaces. (a) Particles 
which sedimented to the bottom surface in a water suspension (inactive system) tend to 
align with their Pt caps facing downwards, which is more pronounced for larger particles. 
The Pt caps ($\rho$ = 21.45 g/cm$^3$), which are much denser than the silica parts of the 
particles, render them bottom-heavy (in the image one sees transparent SiO$_2$ on top of 
heavily absorbing Pt). However, on addition of H$_2$O$_2$ (active system), the particles 
reorient their symmetry axis parallel to the bottom surface and can be seen as half 
covered circles in the right micrograph in (a), where dark semi-spheres correspond to the 
Pt cap, and the SiO$_2$ parts which do not absorb the light appear lighter. (b) Schematic 
of a particle rotating from bottom-down configuration upon peroxide addition. 
(c) Micrograph of Janus colloids ($R = 2.5~\mathrm{\mu m}$) in the vicinity of the step 
with height $h_{step} = 800~\mathrm{nm}$. In the absence of H$_2$O$_2$ (left image), the 
step (seen as a black line) has no influence on the orientation of the particles, (their 
caps are facing downwards, same as far from the step). Upon addition of fuel (right 
image), the particles orient with their symmetry axis parallel to both the bottom surface 
and the step. Scale bars in (a) and (c) correspond to 5 $\mathrm{\mu m}$. (d) The phase 
portrait for a bottom-heavy (see Supplementary Note 1) particle with $R = 2.5~\mathrm{\mu 
m}$ at an infinite planar wall oriented with its normal parallel to the direction of 
gravity. The phase portrait is calculated at $b_{inert}/b_{cap} = 0.3$, $b_{w}/b_{cap} 
= -0.2$, where $b_{inert}$ and $b_{cap}$ are the surface mobilities (see Methods for 
details) at the inert and catalytic faces of the particle respectively, and $b_{w}$ 
is the surface mobility at the wall. (The phase portrait for $R = 1.0~\mathrm{\mu m}$ 
is shown in Supplementary Fig. 2) The phase portrait indicates that a particle will 
rotate to its steady state orientation $\theta = \theta_{eq} \approx 90^\circ$ for all 
initial conditions. The inset represents a schematic diagram of the system: a Janus sphere 
of radius $R$ is placed at distance $h$ above an inert wall; $\theta$ describes the 
orientation of the particle’s cap with respect to the wall normal. $\Delta x$ is 
the distance from the step to the particle center and $\phi$ is the cap orientation with respect to the step normal. 
(e) The rate of rotation $\dot \theta = − \Omega_x/\Omega_0$ of a particle with $R = 
2.5~\mathrm{\mu m}$ above a planar substrate, including contributions from activity,
gravity, and chemi-osmotic flows on the substrate. This function is the sum of panels 
(a), (e), and (f) in Fig. \ref{fig3}. (f) Phase portrait similar to the one in (d) but 
in absence of gravity; all other parameters are as in (d). This portrait is supposed to 
capture qualitatively the effect of the vertical step wall.
}
\end{figure*}
Initially, the particles are introduced to the system with no H$_2$O$_2$ present, 
and due 
to their weight they sediment near the bottom surface. After sedimentation, we find the 
particles uniformly distributed over the substrate, and most of the 5 $\mathrm{\mu 
m}$ particles have their much denser (compared with the SiO$_2$ cores) Pt-caps oriented 
downwards (Fig. \ref{fig2}a, left), while smaller particles have a wider distribution of
orientations. The particles are seen in the same focal plane of the microscope, which 
indicates that they are at similar vertical distances from the substrate. Upon addition 
of H$_2$O$_2$ to the system, we observe that the Pt-caps of the microswimmers are 
oriented parallel to the substrate plane (see Fig. \ref{fig2}a, right and b; and for the 
definition of the geometrical parameters see Supplementary Figure 1). Following this 
re-orientation, the microswimmers start moving parallel to the substrate in the 
direction away from the catalytic caps. In Fig. \ref{fig2}c we show snapshots from an 
optical microscopy video recording of Janus microswimmers in the vicinity of a step 
with $h_{step} = 800~\mathrm{nm}$. Similarly to the case depicted in Fig. \ref{fig2}a
(left), in the absence of H$_2$O$_2$ the particles are oriented cap-down. After addition of 
hydrogen peroxide, the particle caps turn away from the substrate and the particles start 
moving in random directions until some of them encounter a step; if the step is 
sufficiently tall (depending on the particle size) the particles stop, reorient, and 
continue self-propelling along it. These observations confirm our hypothesis that the 
presence of a side step near the active microswimmers, even if small compared to the 
particle radius, has an influence on their orientation.

We show that this behavior (alignment with both the wall and the step) is captured by a simple 
model of neutral self-diffusiophoresis (see Methods section for details of the model), in 
which we assume that the activity of the Janus particle is captured by the release of a 
neutral solute (O$_2$ molecules) at a constant rate from its catalytic cap \cite{ref28}. 
The resulting anisotropic solute distribution around the particle drives a surface flow 
in a thin layer surrounding the particle, leading to its directed motion 
\cite{ref28,ref29}. The catalytically active particle has several types of interaction 
with a nearby impermeable wall. The particle drives long range flows in the suspending 
solution. These flows are reflected from the wall, coupling back to the particle 
(``hydrodynamic interaction''). Secondly, the particle's self-generated solute gradient 
is modified by the presence of the wall. The wall-induced modification of the solute 
concentration field can contribute to translation and rotation of the particle 
(``phoretic interaction''). In particular, when the solute interacts more weakly with the 
inert region of the particle than with the catalytic cap and both interactions are 
repulsive (see Methods for details), the confinement and accumulation of solute near
the substrate tends to drive rotation of the cap away from the substrate. On the other 
hand, the bottom-heaviness of the particle, along with the hydrodynamic interaction of 
the particle with the substrate, tends to drive the rotation of the cap toward the 
substrate. Finally, the inhomogeneous solute distribution along the wall induces a solute 
gradient driven ``chemi-osmotic'' flow along the substrate. 
For repulsive solute-substrate 
interactions, this surface slip velocity is directed quasi-radially inward toward the 
particle, driving a particle-uplifting flow in the suspending solution, as well as causing
the particle cap to rotate away from the substrate. For attractive solute-substrate 
interactions the opposite directions of flows apply. Numerical analysis of this model 
system shows that, depending on the relative strengths of these interactions (i.e. the 
parameters characterizing the surface chemistry of the particle and the wall), the 
various contributions to rotation discussed above may balance at a steady height 
$h_{eq}$ and orientation $\theta_{eq} \approx 90^\circ$, and that this steady state is 
robust and stable against perturbations in height and orientation. The particle cap 
orientation would therefore evolve to $\theta_{eq} \approx 90^\circ$ (i.e., the symmetry 
axis almost parallel to the substrate) from nearly all initial orientations, including a 
cap-down one. In Fig. \ref{fig2}d, a phase portrait shows the dynamical evolution of 
particle height and orientation, and the color-coded rate of rotation $\dot \theta = 
−\Omega_x$ is depicted in Fig. \ref{fig2}e. The steady state (red dot) clearly has a 
large basin of attraction. We note that our numerical calculations were carried out
for $h/R \geq 1.02$. Therefore, some trajectories in the region of the cap up ($\theta = 
180^\circ$) orientation encounter a numerical cutoff. However, based on the structure of 
the phase portrait, we expect such trajectories to roll toward $\theta \approx 
90^\circ$ after close encounter with the wall.
\begin{figure*}[!bht]
\includegraphics[width = 0.95\textwidth]{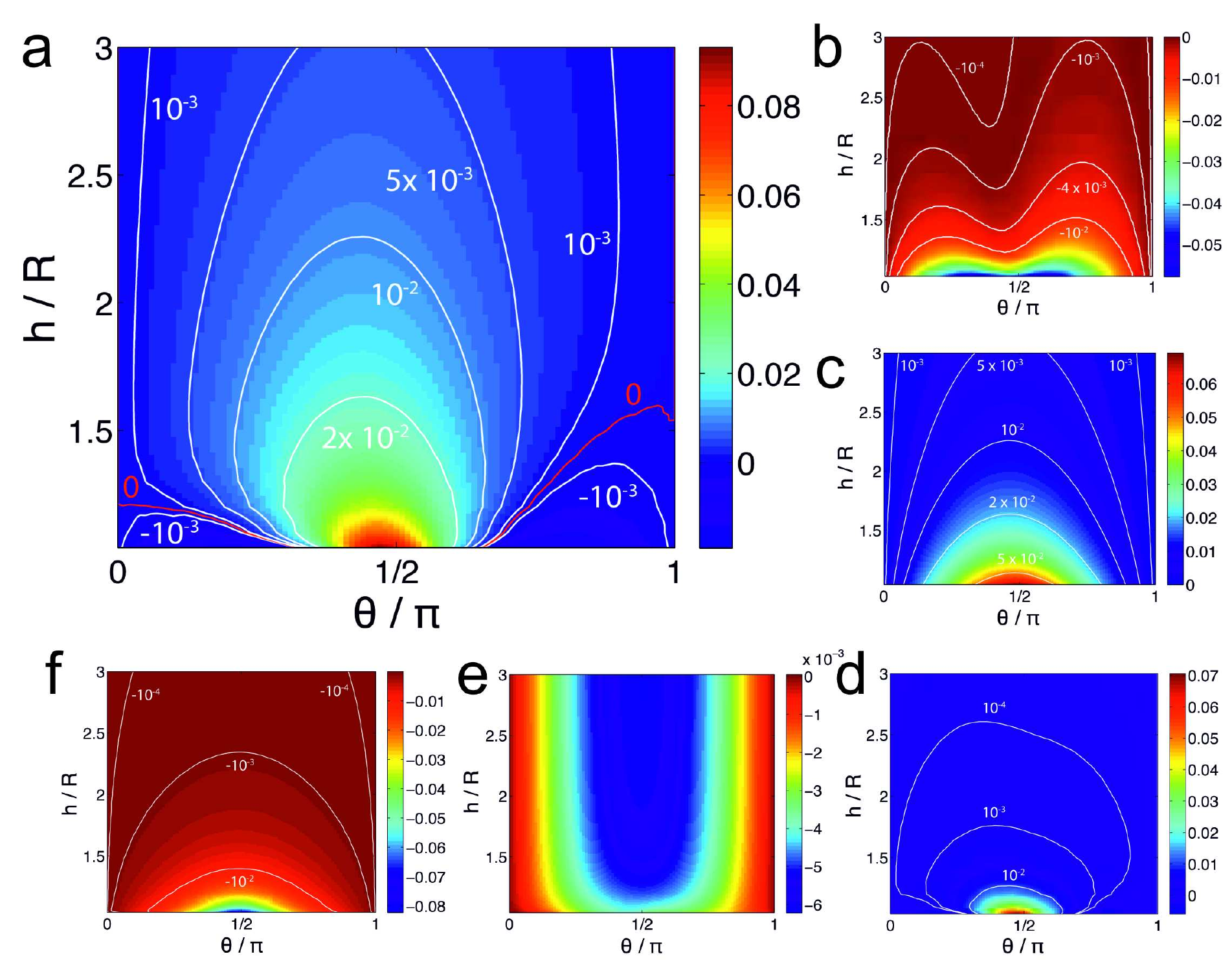}
\caption{\label{fig3} Various contributions to particle angular velocity. (a) 
Contribution from self-diffusiophoresis $-\Omega_x^a/\Omega_0$ (see Supplementary Note 2 
and Eq. (2) therein) as a function of height $h/R$ and orientation $\theta$ for half 
covered Janus microswimmer and unequal surface mobilities $b_{inert}/b_{cap} = 0.3$. 
Throughout, white curves correspond to constant values of $\Omega_x^a$. Note that, by 
definition, panel (a) is the sum of panels (b), (c), and (d). (b) Contribution 
$-\Omega_x^{a,hi}/\Omega_0$ obtained by using the free space number density of solute 
distribution $c^{fs}(\mathbf{r})$ around the particle, i.e., neglecting the influence of 
the wall on the number density of solute, but including the influence of the wall on the 
hydrodynamic flow. (c) Contribution $-\Omega_x^{a,sol}/\Omega_0$ obtained by using the 
free space hydrodynamics stress tensor $\boldsymbol{\sigma'}^{fs}$ in the dual Stokes 
problems employed in the reciprocal theorem, i.e., neglecting the effect of the wall on 
the hydrodynamics, but including the chemical effect. (d) Contribution 
$-\Omega_x^{\delta \delta}/\Omega_0$ due to higher order coupling between the two 
effects. (e) Contribution $-\Omega_x^{a,g}/\Omega_0$ to rate of rotation from the 
bottom-heaviness of the particle. (f) Chemio-osmotic contribution 
$-\Omega_x^{a,ws}/\Omega_0$ due to the activity-induced phoretic slip at the wall 
calculated at $b_{w}/b_{cap} = -0.2$.
}
\end{figure*}

If a particle as above would encounter now a second vertical side wall, 
numerical simulations for the same interaction parameters show (see Fig. \ref{fig2}f) 
that for this wall, for which gravity now plays no role,  a similar sliding along the 
wall attractor emerges with $\phi_{eq} \approx 90^\circ$, i.e. with the particle 
oriented with its axis almost parallel to the vertical wall. The combination of the two 
sliding states thus aligns the axis of the particle along the edge formed by the two 
walls. Note that although this second fixed point appears to have a smaller basin of 
attraction, it should capture the whole $\phi \leq 90^\circ$ range. A particle on a 
trajectory that ``crashes'' into the vertical wall would diffuse along the wall until it 
reaches the basin of attraction in the vicinity of $\phi_{eq} \approx 90^\circ$. While 
the argument is developed for the superposition of two infinite planar walls\footnote{The 
above results will also hold for particles with the catalytic cap less dense than Pt. It 
is easy to see that in this case (while keeping all the other parameters of the system 
like geometry of the cap, activity, etc., fixed) a sliding fixed point along the bottom 
wall will also emerge. Moreover, we have checked via numerical simulations (results not 
shown) that the corresponding height and the orientation will lie between the values 
corresponding to the fixed points shown in Figs. \ref{fig2}d and \ref{fig2}f. Thus the corresponding orientation will remain close to $90^\circ$.}, we expect that similar features may occur for a vertical step with finite height.

Within our model, we can isolate and quantify the various wall-induced contributions to 
particle motion discussed above. The mathematical details of the decomposition are given 
in Supplementary Note 2. In Fig. \ref{fig3}b we show the contribution to the rate of 
rotation $\dot \theta$ of the particle from hydrodynamic interaction (HI) with the wall 
as a function of particle height and orientation. Hydrodynamic interactions always rotate 
the particle cap towards the wall. Therefore, for the particular combination of parameters
used in this work, hydrodynamic interactions cannot by themselves produce a steady 
orientation $\theta_{eq} \approx 90^\circ$. On the other hand, phoretic interactions 
always rotate the cap away from the wall, as described above (Fig. \ref{fig3}c). 
Therefore, the interplay of hydrodynamic and phoretic interactions can produce a curve 
with $\dot \theta = 0$ in the region of $\theta_{eq} \approx 90^\circ$ (Fig. 
\ref{fig3}a). Moreover, the contributions of bottom-heaviness (Fig. \ref{fig3}e) and 
chemi-osmotic flow on the wall (Fig. \ref{fig3}f) to the angular velocity are comparable 
in magnitude to the contributions from hydrodynamic and phoretic interactions. Therefore, 
for the parameters used in this work, all of these effects are important in determining 
the emergence and location of a ``sliding state'' attractor. The surface chemistry 
parameters were chosen as providing the best fit to the experimental observations of the 
two sliding states (above a substrate and along a side wall).
\begin{table*}[!bht]
\begin{center}
  \begin{tabular}{ c | c | c | c | c | c}
    \hline
    \textbf{Model} & \textbf{Best fit parameters} & $h_{eq}/R$, \textbf{no 
gravity} & $\theta_{eq}$, \textbf{no gravity} & $h_{eq}/R$, \textbf{with gravity} & 
$\theta_{eq}$, \textbf{with gravity} \\ \hline \hline
  & $b_{inert}/b_{cap} = 0.3$ &  &  &  &  \\ 
Full Model  & $b_{wall}/b_{cap} = -0.2$ & 1.11 & $77.9^\circ$ & 1.06 & $94.8^\circ$ \\
  & $b_{cap} < 0$ &  &  &  &  \\   
    \hline
Squirmer, first  &  &  &  &  &  \\
two squirming & $B_2/B_1 = 0.3$ & 1.64 & $102^\circ$ & $< 1.02$ & $\approx 45^\circ$ \\
modes only &  &  &  &  &  \\ \hline
Effective & $b_{inert}/b_{cap} = -0.8$ &  &  &  &  \\
squirmer & $b_{cap} < 0$ & 1.063 & $69.7^\circ$ & 1.09 & $65.3^\circ$ 
\\ \hline
  \end{tabular}
\caption{\label{table1} Comparison of full model with two hydrodynamics-only models. For 
each model, we list the parameters that give the best fit to the experimental 
observations. For each model and set of best fit parameters, we give the height and 
orientation of the particle when it is in a ``sliding state'' above a planar wall in both 
the presence of gravity (corresponding to motion above a substrate) and the absence
of gravity (corresponding to motion near a side wall). Experimentally, it is observed that
$\theta_{eq} \approx 90^\circ$ in both cases. Of the three models, the full model shows 
the best fit with these experimental observations. For the squirmer with only the first 
two squirming modes, there are clear signs of an attractor with $h_{eq}/R$ below the 
numerical cut-off of $h/R = 1.02$ in the presence of gravity, but this attractor has 
$\theta_{eq}$ far from $90^\circ$ (see also Supplementary Note 3 and Supplementary Table 
1). The best fit effective squirmer agrees moderately well with the experimental 
observations. However, the orientation of the sliding seems significantly different from 
the experiment, and, as discussed in Supplementary Note 3 and Supplementary Table 2, 
the best fit parameters correspond to an unrealistically large force dipole.
}
\end{center}
\end{table*}

As noted above our model includes several types of interaction of the particle with the 
wall. However, many theoretical \cite{ref24} and experimental \cite{ref22,ref23} studies 
have sought to characterize the interaction of active particles and solid boundaries 
strictly in terms of effective hydrodynamic interactions (HI). It is therefore 
interesting to compare our full model against the best fit results from effective HI 
models. We consider two such approaches, the details of which are given in Supplementary 
Note 3. Briefly, in the first approach, we use the classical ``squirmer'' model, and 
specify a priori the amplitude of the first two squirming modes. Higher order modes are 
taken to have zero amplitude. In the second approach, we consider the ``effective 
squirmer'' obtained within our model by neglecting phoretic and chemi-osmotic effects. 
The ``effective squirmer'' approach intrinsically covers a broad range of squirming mode
amplitudes. In Table \ref{table1}, we show that the results of the full model match the 
experimental observations significantly better than the best results of the two HI-only 
approaches.

\subsection{Dynamics of Janus microswimmers at a rectangular step}
We designed a system with microfabricated 3D structures by patterning of photoresist 
through a circular or square mask, followed by e-beam deposition of the required material (Si or SiO$_2$ in our case), and then removal of the developed photoresist resulting in desired structures (for detailed information see Methods). Depending on the use of 
positive or negative photoresist we obtain patterns with posts or wells of different 
shapes (see Supplementary Figure 5). The height of the features patterned on a substrate 
is tunable in a wide range; in this study, we have tested step heights $h_{step}$ between 
100 and 1000 nm.

\subsubsection{Characterization of particle trajectories approaching a step}
\begin{figure*}[!bht]
\includegraphics[width = 0.9\textwidth]{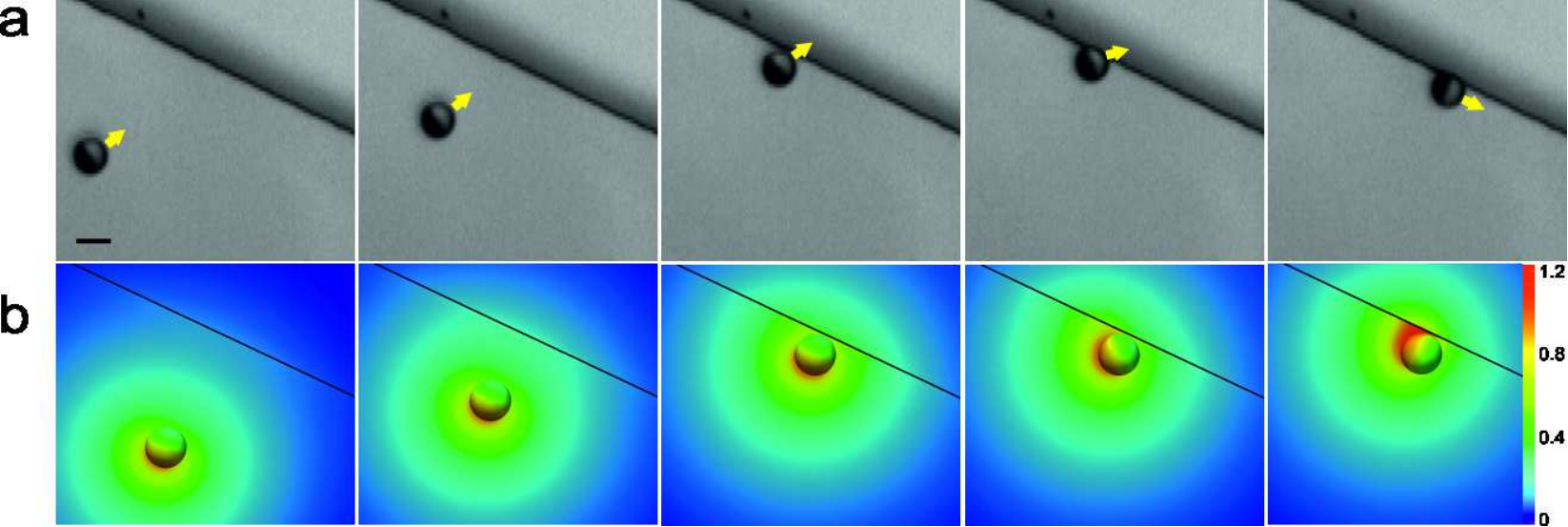}
\caption{\label{fig4} Effect of 800 nm height step on the dynamics of a Janus 
microswimmer. (a) An active Janus particle approaching a step; after direct contact with 
the step, it reorients until its propulsion axis is parallel to the step. $R = 
2.5~\mathrm{\mu m}$, $h_{step} = 800~\mathrm{nm}$, 2.5 \% vol. H$_2$O$_2$. (b) 
Numerically calculated steady-state distribution $c(\mathbf{r})$ of reaction products 
around half catalyst-covered Janus sphere as a function of the step distance and the cap 
orientation with respect to the step. The color map shows $c(\mathbf{r})$ at the 
surfaces of the particle and substrate; $c(\mathbf{r})$ is in units of $c_0$ (see 
Methods). Scale bar correspond to 5 $\mathrm{\mu m}$.
}
\end{figure*}
Fig. \ref{fig4}a shows snapshots of a typical trajectory 
of a microswimmer moving toward a step at almost perpendicular direction. Once the 
particle hits the step (Fig. \ref{fig4}a, third panel) it starts reorienting its axis 
(Fig. \ref{fig4}a, forth panel) toward the direction along the step (Fig. \ref{fig4}a, 
fifth panel). We observe that in most cases the complete process of reorientation takes 
less than ten seconds, independent of the initial angle at which the particle approaches 
the step. Within the resolution of our experimental equipment, we do not observe any 
systematic deflection in the trajectory of the particle in the vicinity of the steps. 
Therefore we conclude that if any long range effective interaction exists between the 
particles and the steps, it must be very weak. This observation is reproduced by our 
numerical model: we calculate that the effects of a wall on the velocity of a particle are 
negligible when the particle is more than three radii away from the wall (Supplementary 
Fig. 4 and Supplementary Note 4). We thus attribute the suppression, upon 
collision with the step, of the motion of the particles normal to the step 
solely to steric interactions.

In Fig. \ref{fig4}b we present the distribution of the reaction products around a 
microswimmer calculated numerically for particle positions and orientations approximately 
corresponding to those shown in the experimental micrographs in Fig. \ref{fig4}a. When 
the particle is far away from the step (Fig. \ref{fig4}b, first and second panels), 
approaching it in a head-on direction, the generated concentration field confirms the
expected mirror symmetry with respect to the plane defined by the motion axis and the 
normal to the substrate. As the result of this symmetry there are no activity induced 
rotations and the particle stays on its head-on track (up to Brownian rotational 
diffusion) toward the step. However, closer to the step a head-on collision becomes 
unstable to small fluctuations of the propulsion axis as any such fluctuation gets 
amplified by the buildup of asymmetric product distribution in the region between one 
side of the particle and the step (Fig. \ref{fig4}b, forth panel). This eventually leads 
to the reorientation of the motion axis parallel to the step (Fig. \ref{fig4}b, fifth 
panel).

\subsubsection{Effects of the step height on the capture efficiency}

\begin{figure}[!htb]
\includegraphics[width = 0.9\columnwidth]{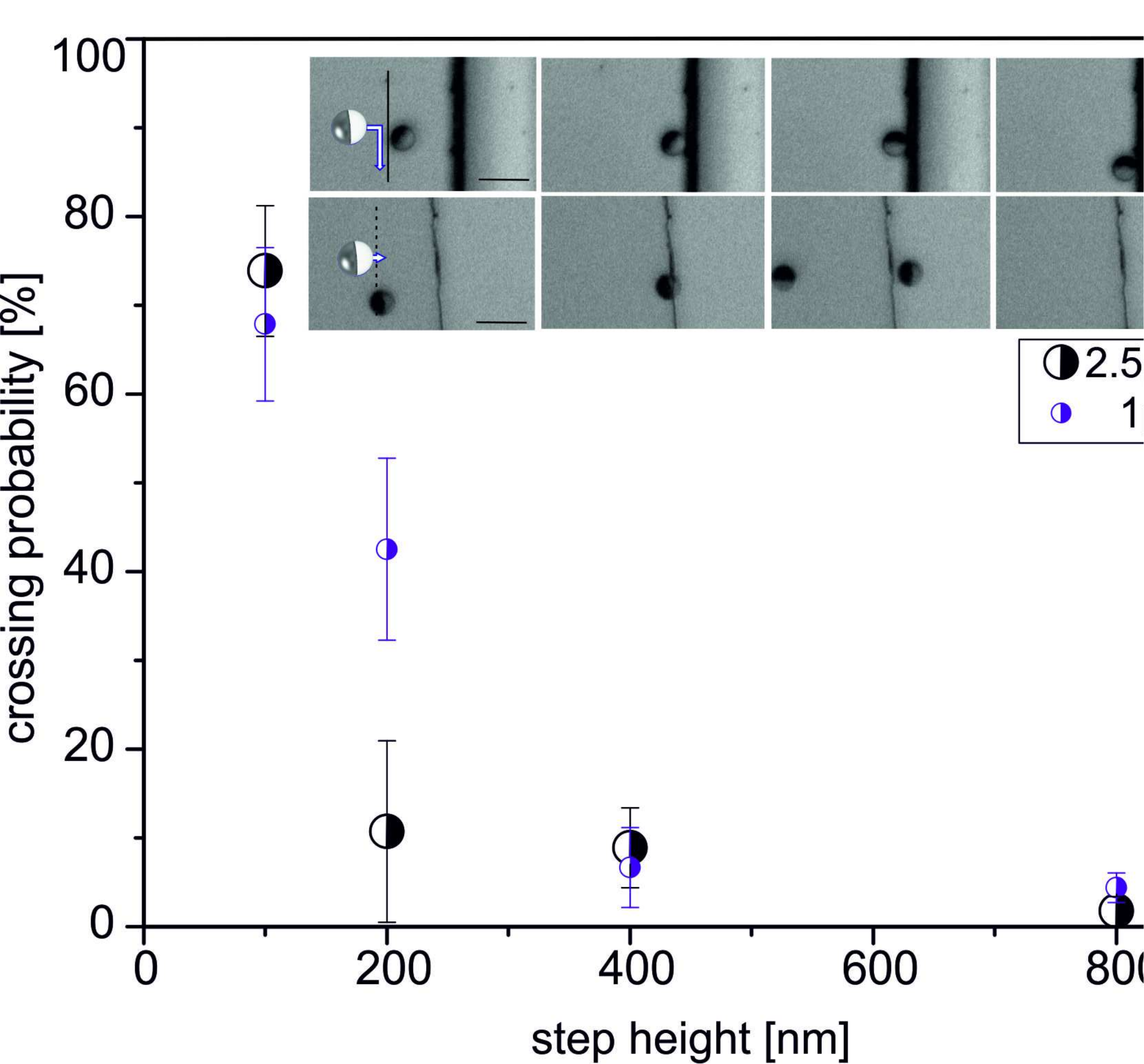}
\caption{\label{fig5} Sub-micron steps as rectifiers of active particles' trajectories. A 
summary of the crossing behavior of Janus SiO$_2$ microswimmers of different sizes at 
several values of $h_{step}$; the error bars are standard errors of the mean. Inset: A 
sequence of micrographs showing a Janus particle with $R = 2.5~\mathrm{\mu m}$ 
approaching a step with $h_{step} = 800$ nm, reorienting and then moving parallel to it. 
Micrograph sequence of a Janus particle, $R = 2.5~\mathrm{\mu m}$, passing over a step, 
$h_{step} = 100$ nm. All scale bars correspond to 10 $\mathrm{\mu m}$.
}
\end{figure}

We observe that submicron steps are able to capture and guide particles as shown in 
Fig. \ref{fig5} (insets). To evaluate the minimum height $h_{step}^{*}$ that can still 
influence the trajectory of the particles, we fabricated a set of patterns with 
$h_{step}$ varying in a range from 100 nm to 1000 nm. The results for the two 
different particle sizes show that $h_{step}^{*}$ decreases as the particle size 
increases. In Fig. \ref{fig5} we summarize the responses of $R = 1.0~\mathrm{\mu m}$
and $R = 2.5~\mathrm{\mu m}$ active particles to steps of different heights. 
Both types 
of particles could swim over the step of 100 nm height. For $R = 2.5~\mathrm{\mu 
m}$ particles steps of height 200 nm already ensure about 90\% docking of particles 
upon collision with the step, while a significant fraction of the $R = 1.0~\mathrm{\mu 
m}$ particles managed to pass over the 200 nm high step, and 400 nm high steps 
were required for efficient docking. From Fig. \ref{fig5} we infer that $h_{step}^{*}/R$
is smaller for larger particles.

Having estimated the threshold values $h_{step}^{*}$ for particle trapping, we now 
select steps of sufficient height to ensure full trapping upon collisions. Therefore, all 
the following experiments were carried out on 800 nm features, for which both 2.5 and 1.0
$\mathrm{\mu m}$ particles follow the step upon collision.

\subsection{Guidance of microswimmers by low height topographic steps}

\begin{figure}[!htb]
\includegraphics[width = 0.9\columnwidth]{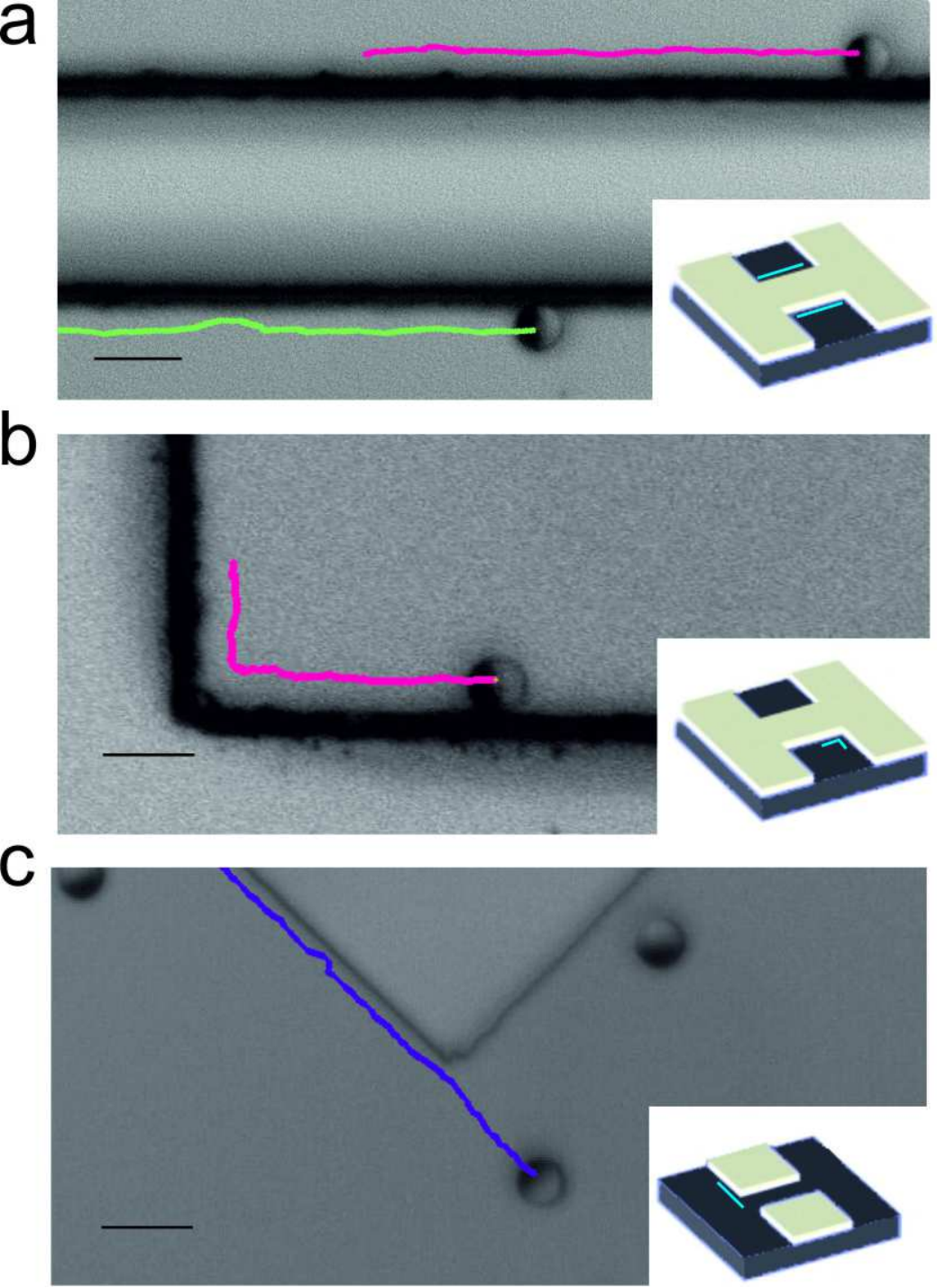}
\caption{\label{fig6} Guidance of Janus microswimmers by step features. (a) A micrograph 
showing trajectories of two $R = 2.5~\mathrm{\mu m}$ Janus particles following a straight 
step (b) A Janus particle tracked while maneuvering around a $90^\circ$ corner. (c) A 
Janus particle unable to follow a reflex angle of $270^\circ$. The insets 
show schematically the structures of wells (a), (b), and posts (c). The blue lines on the 
insets schematically indicate the position of Janus particles in actual experiments. 
Scale bars correspond to 10 $\mathrm{\mu m}$.
}
\end{figure}

The step-like topography and particle alignment along step edges can be used to guide 
microswimmers, as it is shown in Fig. \ref{fig6}a. 
\begin{figure*}[!thb]
\includegraphics[width = 0.85\textwidth]{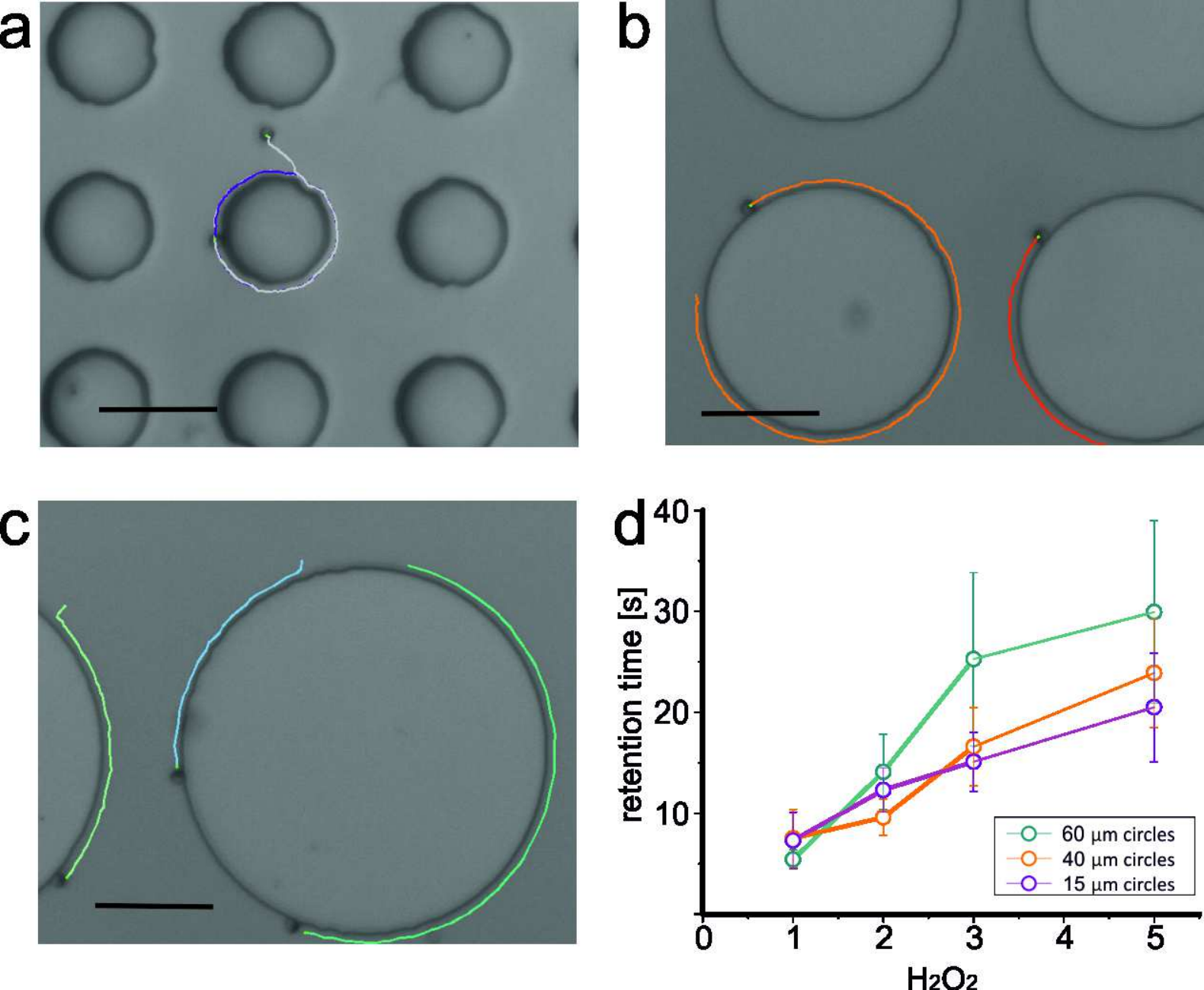}
\caption{\label{fig7} Trapping and guiding active particles by circular posts. (a–c) Optical 
snapshots of $R = 1.0~\mathrm{\mu m}$ active particles moving for 12 s around circular 
posts of 15, 40 and 60 $\mathrm{\mu m}$ diameter $d$, respectively. Scale bar, 20 $\mathrm{\mu m}$. 
(d) Average retention time as a function of peroxide concentration for $R = 1.0~\mathrm{\mu m}$. 
The average is determined for 15 -- 20 trajectories per data point; the error bars are standard 
error of the mean.}
\end{figure*}
This corresponds to particles in a 
well structure with straight steps. Upon collisions, the particles align along the steps 
and follow them (Fig. \ref{fig6}a and Supplementary Movie 1). In the same well-like 
structure, particles eventually encounter a corner and after spending some time
adjusting their orientation can maneuver around the corner (Fig. \ref{fig6}b, 
Supplementary Movie 2). In the case of a post-like structure as displayed in Fig. 
\ref{fig6}c, particles also follow straight features, but fail to reorient and maneuver 
around the $270^\circ$ corner (see also Supplementary Movie 3). These findings
suggest that certain critical value must exist for reflex angles between features above which guidance along the edge of the post is lost.

Active Janus microswimmers also follow circular trajectories around circular posts as 
shown in Fig. \ref{fig7}, which requires constant reorientation of the axis. In Fig. 
\ref{fig7}a-c paths are shown where particles with $R = 1.0~\mathrm{\mu m}$ circle 
around posts with the diameter of 15, 40 and 60 $\mathrm{\mu m}$, respectively, for more 
than twelve seconds (see Supplementary Movies 4-6). Supplementary Fig. 3 and Movie 7 show 
an additional example of cycling motion with long retention times. We find that the 
retention time of microswimmers at the circular posts increases with increasing peroxide 
concentration, as displayed in Fig. \ref{fig7}d. At 1\% H$_2$O$_2$, few 
particles completely circle around a whole post and in most cases the microswimmers 
detach from the post before a complete revolution (at lower peroxide concentrations
the particles hardly move and get easily stuck at the steps, so measurements were not 
considered). At 2\% H$_2$O$_2$ the path length along the posts is increased, and 
likewise in 3\% and 5\% H$_2$O$_2$, where many particles circle around posts multiple 
times. At even higher concentrations of H$_2$O$_2$, we observe vigorous formation of 
oxygen bubbles and occurrence of convective flows; thus no reliable
measurements could be performed above 5\% H$_2$O$_2$.

Since the retention time increases with the concentration of H$_2$O$_2$, we 
conclude that it is the activity of the microswimmers which is directly responsible for 
the effective particle attraction to the posts as well as for the occurrence of the 
sliding attractor: it is the net result of the particle-step hydrodynamic interaction and 
confinement induced modification of the distribution of the solute concentration. The
strength of both effects depends on the fuel concentration: increased fuel concentration 
leads to a higher production rate of solute (i.e. stronger phoretic and chemi-osmotic 
interactions) and a higher self-propulsion velocity (i.e. stronger hydrodynamic 
interactions). The finite retention time is set by the competition between 
activity-induced effective attraction to the post side walls and the rotational
diffusion. Increased fuel concentration increases the strength of the first factor without 
affecting the second one, and therefore increases the retention time. The robustness of 
the sliding state attractor is further discussed in Supplementary Note 5.

\section{Discussion}
 
We report experimental results showing the dynamics of chemically active Janus 
microswimmers at geometrically patterned substrates and a qualitative interpretation in 
terms of a minimal continuum model of self-diffusiophoresis of chemically active 
colloids. Employing a lithography-based method to fabricate submicron topographic 
features in the form of rectangular stripes, square posts, cylindrical posts or square 
wells on glass surface or silicon wafer, we demonstrate that the motion of chemically
active Janus microswimmers can be restricted to proceed along these small height patterns 
for significant time intervals. Furthermore, the motion along the circumference of 
cylindrical posts reveals that the retention time increases with increasing 
H$_2$O$_2$ concentration. This allows us to unequivocally identify the particle’s 
chemical activity, which modulates the distribution of the phoretic slip at the
particle surface and thus the hydrodynamic interaction with the nearby topography, as 
playing a dominant role in the observed phenomenology.

We also show that a minimalist, continuum model of self-diffusiophoresis captures 
the qualitative features of the experimental observations if 
one accounts for the difference in material properties of the two parts of the colloid, 
as well as for chemi-osmotic flows induced at the wall. This latter aspect highlights the 
need for models that explicitly include chemical activity, without which a no-slip 
boundary condition would apply at the wall. The model employed here allows us to understand the 
emergence of states of motion along the edges as a simultaneous attraction to two
fixed-point attractors corresponding to steady sliding states along the bottom wall and 
along the vertical wall of the step.

The micro-structuring method presented here avoids the use of any external fields and 
relies solely on the intrinsic properties of the system to control particle motion. The 
phenomenology reported here is, in some sense, a mesoscale analogue of the binding of 
motor-proteins to microtubules to switch to directional motion. However, in distinction 
to biological nanomotors, the Janus microswimmers bypass the binding and rather elegantly 
exploit an effective attraction that stems from the feedback between geometric 
confinement and chemical and hydrodynamic activity. The results presented here open the
possibility of robust guidance of particles along complex paths via minimal surface 
modifications, i.e., by sculpting a pattern with the edge in the desired shape. This may 
have significant implications in designing new applications based on artificial swimmers. 
Finally, we consider that these findings will allow further developments by employing 
smart, chemically patterned walls, where features of the nearby surfaces (and thus the guiding of the microswimmers) can be turned on and off.

\section{Methods}

\subsection{Sample preparation}

Janus particles were obtained by drop casting of a suspension of spherical silica colloids 
(diameter of 2 or 5 $\mathrm{\mu m}$, Sigma Aldrich) on an oxygen-plasma cleaned glass 
slide followed by slow evaporation of the solvent and subsequent placement in an 
e-beam system. High 
vacuum was applied and subsequently a monolayer of 7 nm Pt was evaporated to guarantee 
catalytic properties. To release particles from the glass slides into deionized water, short ultrasound pulses were sufficient.

Photoresist patterns were prepared on 24 mm square glass slides or with the same method on 
silicon wafers. In case of positive photoresist AR-P 3510 was spin-coated onto the 
cleaned substrate at 3500 rpm for 35 s, followed by a soft bake using a hotplate at 90 
$^\circ$C for 3 min and exposure to UV light with a Mask Aligner (400 nm) for 2 s. 
Patterns were developed in a 1:1 AR300-35:H$_2$O solution. In case of negative 
photoresist a layer of TI prime was spin-coated on the substrate during 20 s at 3500 rpm.
After 2 min of drying at 120 $^\circ$C the negative photoresist was coated employing a 
program of 35 s spinning at 4500 rpm, followed by 5 min baking at 90 $^\circ$C. The 
exposure was carried out with a Mask aligner for 2 s followed by 2 min on the hotplate at 
120 $^\circ$C. Finally an additional exposure to 2 s UV light is applied and the patterns 
were developed in pure AZ726MIF. The steps were obtained by e-beam deposition the
desired material (SiO$_2$, Si) in the desired thickness. By dissolving the photoresist 
layer in Acetone the pattern structures of the substrate are exposed, the whole process 
is illustrated in Supplementary Figure 5.

Prior to experiment the patterned substrates were cleaned by oxygen plasma. Experiments 
were performed directly on the substrates by adding equal volumes of particles in DI 
water and diluted peroxide solutions. Videos were recorded with a Leica DFC 300G camera 
mounted to a Leica upright microscope at approx. 30 fps. Evaluation and tracking was 
performed using Fiji analysis software.

\subsection{Tracking}
Accurate tracking of Janus particles was performed automatically by a specially developed 
script in Python 2.7 using the OpenCV library. The position of the Janus swimmers at 
every frame is found by extracting the background, which erases the static posts from the 
image, leaving only the moving particles.

\subsection{Theoretical Modeling}

We model particle motion within a continuum, neutral self-diffusiophoretic framework. A 
particle emits solute at a constant rate from its catalytic cap. The number density 
$c(\mathbf{r})$ of solute, where $\mathbf{r}$ is a position in the fluid, is quasi-static. 
The solute field is governed by the Laplace equation $\nabla^2 c = 0$, and obeys the 
boundary conditions $- D \nabla c \cdot \mathbf{n} = \kappa$ on the catalytic cap 
and $- D \nabla c \cdot \mathbf{n} = 0$ on the inert face of the particle and the 
substrate, where $\kappa$ is the rate of emission (uniform over the cap), $D$ is the 
diffusion coefficient of oxygen, and $\mathbf{n}$ is the local surface normal. Our model 
neglects the details of the catalytic reaction, which might involve the transport of 
charged intermediates \cite{ref28,ref29}. Nevertheless, we expect this model to capture 
the gross effects of both near-wall confinement of the solute field and hydrodynamic
interaction with nearby walls. The surface gradient of solute drives a surface flow 
(``slip velocity'') in a thin fluid layer surrounding the particle surface 
$\mathbf{v}_s(\mathbf{r}) = - b_s(\mathbf{r}) \nabla_{||} c$, where $\nabla_{||}$ denotes 
the projection of the gradient operator along the surface of the particle.

The coefficient $b(\mathbf{r})$ of the slip velocity, the so-called ``surface mobility'', 
is determined by the molecular interaction potential between the solute and the particle 
surface \cite{ref30}. We allow $b(\mathbf{r})$ to differ between the inert and catalytic 
regions, but assume it is uniform in each region, i.e. take $b_s = b_{inert}$ or $b_s = 
b_{cap}$. Additionally, when we consider the effect of chemi-osmotic flow on the 
substrate, we calculate a wall slip velocity $\mathbf{v}_w(\mathbf{r}) = -b_w \nabla_{||} 
c$, where $b_w$ is a constant. We always take the interaction between the solute and 
particle surface to be repulsive, i.e. $b_s < 0$, so that the model is consistent with the 
observed motion of particles away from their caps.

The velocity $\mathbf{u}(\mathbf{r})$ in the fluid is governed by the Stokes equation 
$-\nabla p + \nabla^2 \mathbf{u} = 0$ and the incompressibility condition $\nabla \cdot 
\mathbf{u} = 0$, where $p(\mathbf{r})$ is the fluid pressure and $\eta$  is the dynamic 
viscosity of the solution. The velocity obeys the boundary conditions $\mathbf{u} = 
\mathbf{v}_w(\mathbf{r})$ on the substrate and $\mathbf{u}(\mathbf{r}) = \mathbf{U}^a + 
\mathbf{\Omega}^a \times(\mathbf{r}-\mathbf{r}_0) + \mathbf{v}_s(\mathbf{r})$ on the 
particle surface, where $\mathbf{r}_0$ is the position of the particle center, and 
$\mathbf{U}^a$ and $\mathbf{\Omega}^a$ are the contributions of activity to the 
translational and rotational velocities of the particle. To obtain $\mathbf{U}^a$ and 
$\mathbf{\Omega}^a$ for a given position and orientation of the particle, we first solve 
for $c(\mathbf{r})$ numerically, using the boundary element method (BEM) \cite{ref31}. 
The slip velocities $\mathbf{v}_s$ and $\mathbf{v}_w$ are then calculated from 
$c(\mathbf{r})$. Inserting the slip velocities in the boundary conditions, and 
requiring that the particle is force and torque free, we solve the Stokes equation 
numerically via the BEM in order to obtain $\mathbf{U}^a$ and $\mathbf{\Omega}^a$ in
terms of characteristic velocity scales $U_0 := |b_{cap}| \kappa/D$ 
and $\Omega_0 := U_0/R$. Additionally, $c(\mathbf{r})$ is calculated
in terms of a characteristic concentration $c_0 := \kappa R/D$.

When we include the effects of gravity, we adopt the geometrical model of Campbell and 
Ebbens, taking the Janus particle as having a platinum cap that smoothly varies in 
thickness between a maximum of 7 nm at the pole and zero thickness at the particle 
equator \cite{ref32}. The gravitational contributions to particle velocity, 
$\mathbf{U}^g$ and $\mathbf{\Omega}^g$, are calculated using standard methods (see 
Supplementary Note 1).

We obtain complete particle trajectories by numerically integrating $\mathbf{U} = 
\mathbf{U}^a + \mathbf{U}^g$ and $\mathbf{\Omega} = \mathbf{\Omega}^a + 
\mathbf{\Omega}^g$. Further details of the numerical method are given in 
Ref. \cite{ref21}. We note that the assumption that the solute field is quasi-static is 
valid in the limit of small P{\'e}clet number $\mathrm{Pe} = U_0 R/D$. We have neglected 
the inertia of the fluid, which is valid for small Reynolds number $\mathrm{Re} = \rho 
U_0 R/\eta$, where $\rho$ is the mass density of the solution. These dimensionless 
numbers are $\mathrm{Pe} \approx 4 \times 10^{-3}$ and $\mathrm{Re} \approx 10^{-5}$ 
for a 5 $\mathrm{\mu m}$ catalytic Janus particle that swims at 6 $\mathrm{\mu m~s^{-1}}$ 
\cite{ref33}.\newline

\noindent \textbf{Author contributions}\newline
S.S. and J.S designed the experiments. J.S. and J.K. performed the experiments and analyzed 
the data. M.T. and W.E.U. performed numerical calculations. M.N.P. contributed to the 
theoretical analysis of the numerical results. J.S., W.E.U., and M.T. wrote the 
manuscript. All the authors discussed the results and commented on the manuscript.\newline

\noindent \textbf{Acknowledgements}\newline
The authors thank Albert Miguel Lopez for help with the automated tracking program. 
W.E.U., M.T., and M.N.P. acknowledge financial support from the DFG, grant no. TA 
959/1-1. S.S, J.S and J.K acknowledge the DFG grant no. S.A 2525/1-1. The research also 
has received funding from the European Research Council under the European Union's 
Seventh Framework Programme (FP7/2007-2013)/ERC grant agreement 311529.

\bibliography{refs1}

\onecolumngrid

\newpage


\section*{{\Large Supplementary Information}}

\vspace*{.2in}

\subsection*{{\large Supplementary Figures}}

\renewcommand\thefigure{\textbf{Supplementary Figure \arabic{figure}}}
\setcounter{figure}{0} 
\renewcommand\thetable{\textbf{Supplementary Table \arabic{table}}}
\setcounter{table}{0} 

\vspace*{.5in}

\begin{figure}[!htb]
\includegraphics[width = 0.6\columnwidth]{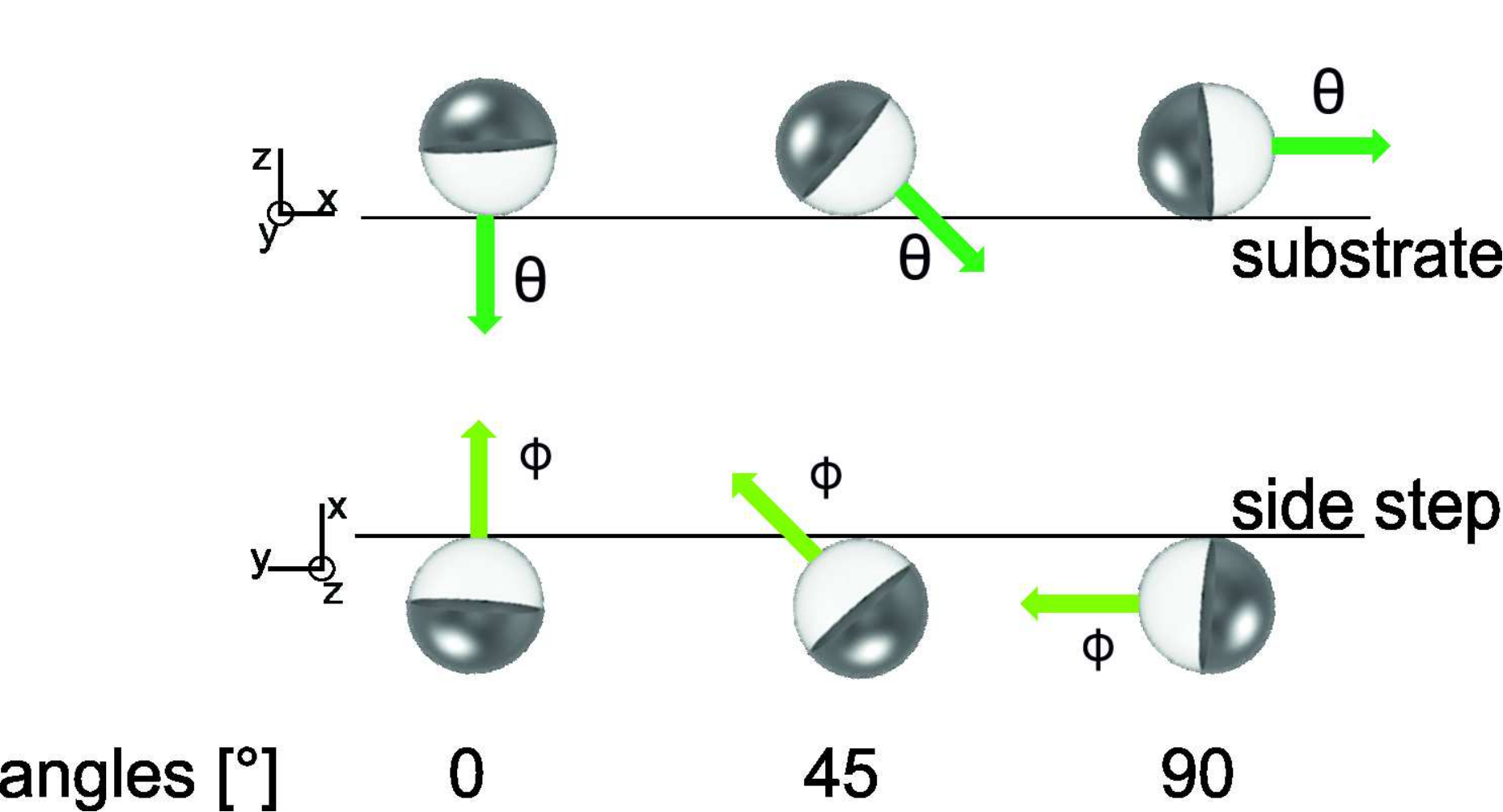}
\caption{\label{SI_fig1} 
Definition of the angles of the 
microswimmer symmetry axis relative to the substrate and the side step.
}
\end{figure}

\vspace*{.5in}

\begin{figure}[!htb]
\includegraphics[width = 0.6\columnwidth]{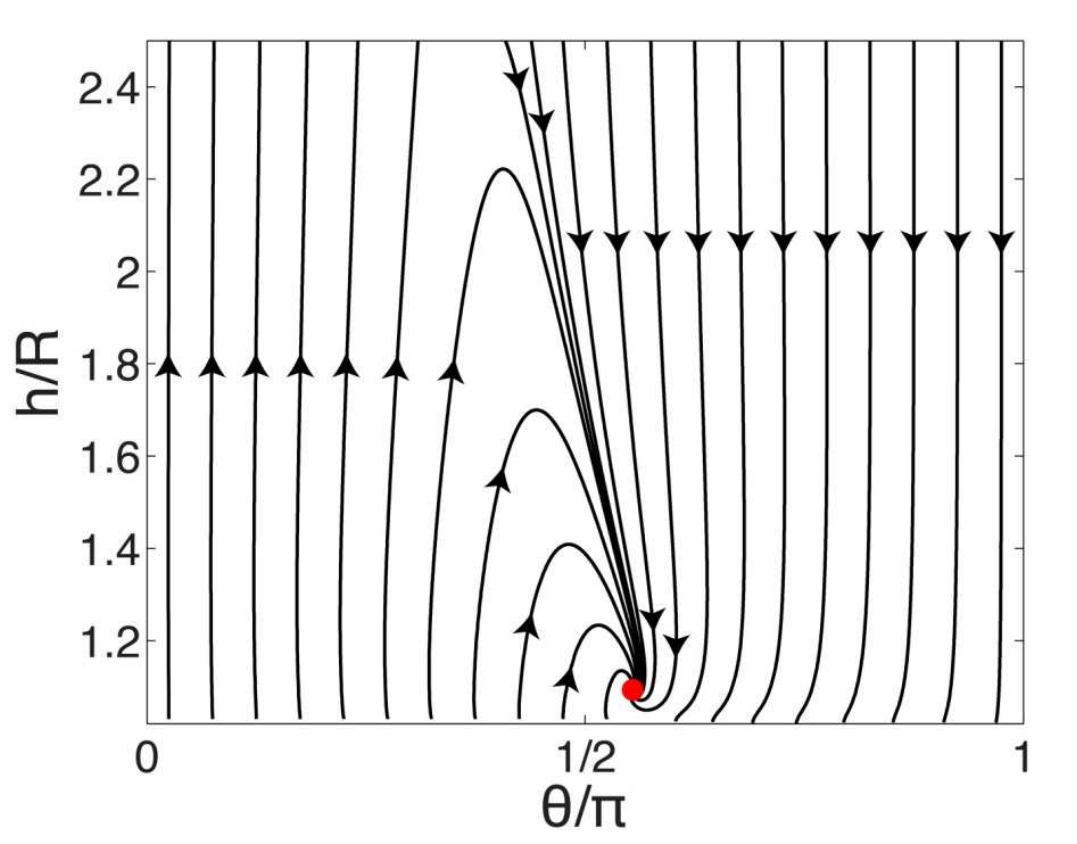}
\caption{\label{SI_fig2} 
Phase portrait for orientation parallel to the substrate 
of a microswimmer with $R = 1.0~\mathrm{\mu m}$, including the effect of gravity. 
Except for the size of the particle, all parameters are the same as in Fig. \ref{fig2}d 
in the main text.
}
\end{figure}

\newpage

\vspace*{0.7 in}

\begin{figure}[!htb]
\includegraphics[width = 0.6\columnwidth]{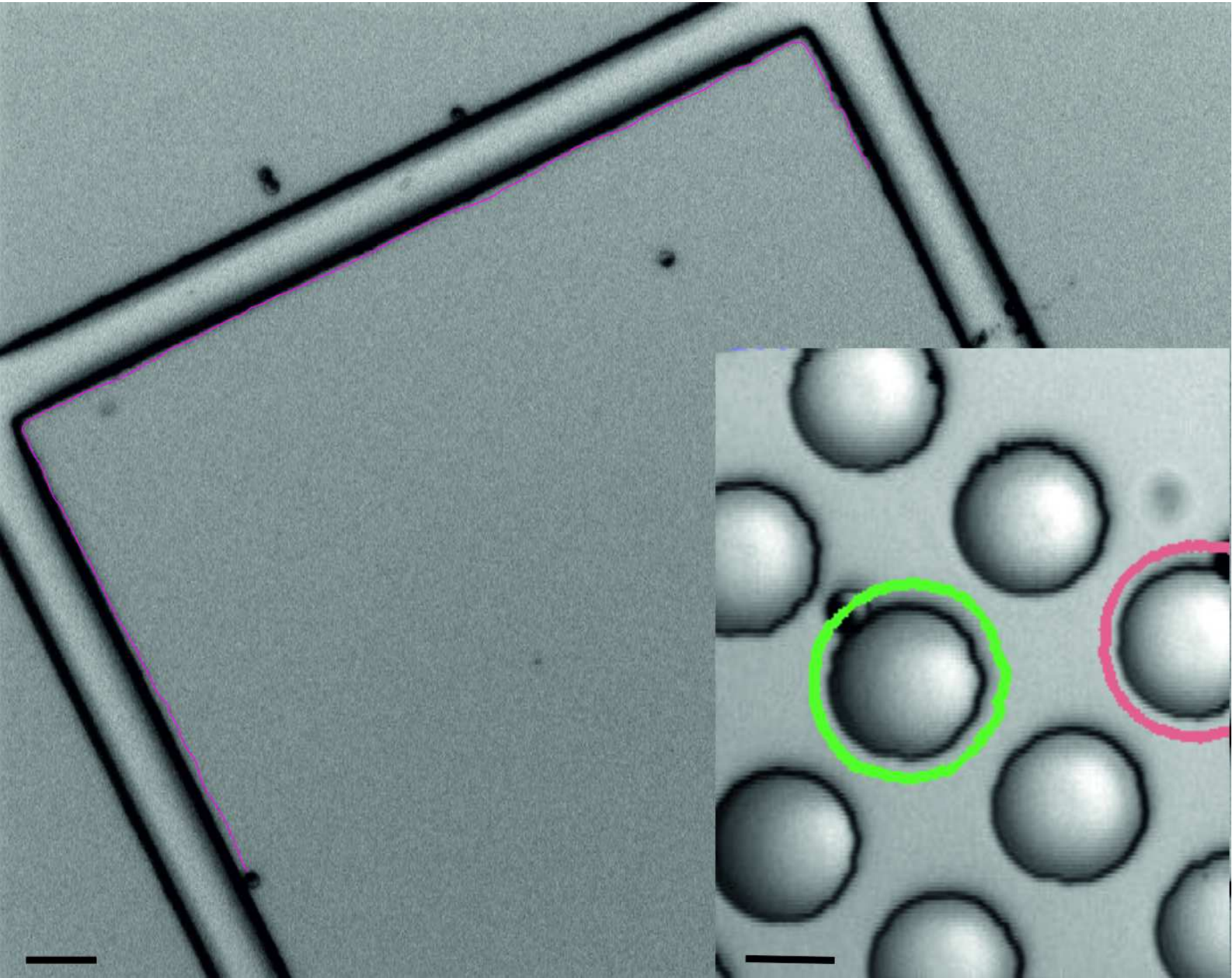}
\caption{\label{SI_fig3} 
$R = 2.5~\mathrm{\mu m}$ Janus microswimmer trajectories 
following well features for 227 s; scale bar corresponds to 20 $\mathrm{\mu m}$. Inset: 
$R = 2.5~\mathrm{\mu m}$ Janus microswimmer circling around a step for more than
89 s; both tracks were recorded in 5 vol\% H$_2$O$_2$, scale bar corresponds to 
10 $\mathrm{\mu m}$.
}
\end{figure}

\vspace*{1.in}

\begin{figure}[!htb]
\includegraphics[width = 0.45\columnwidth]{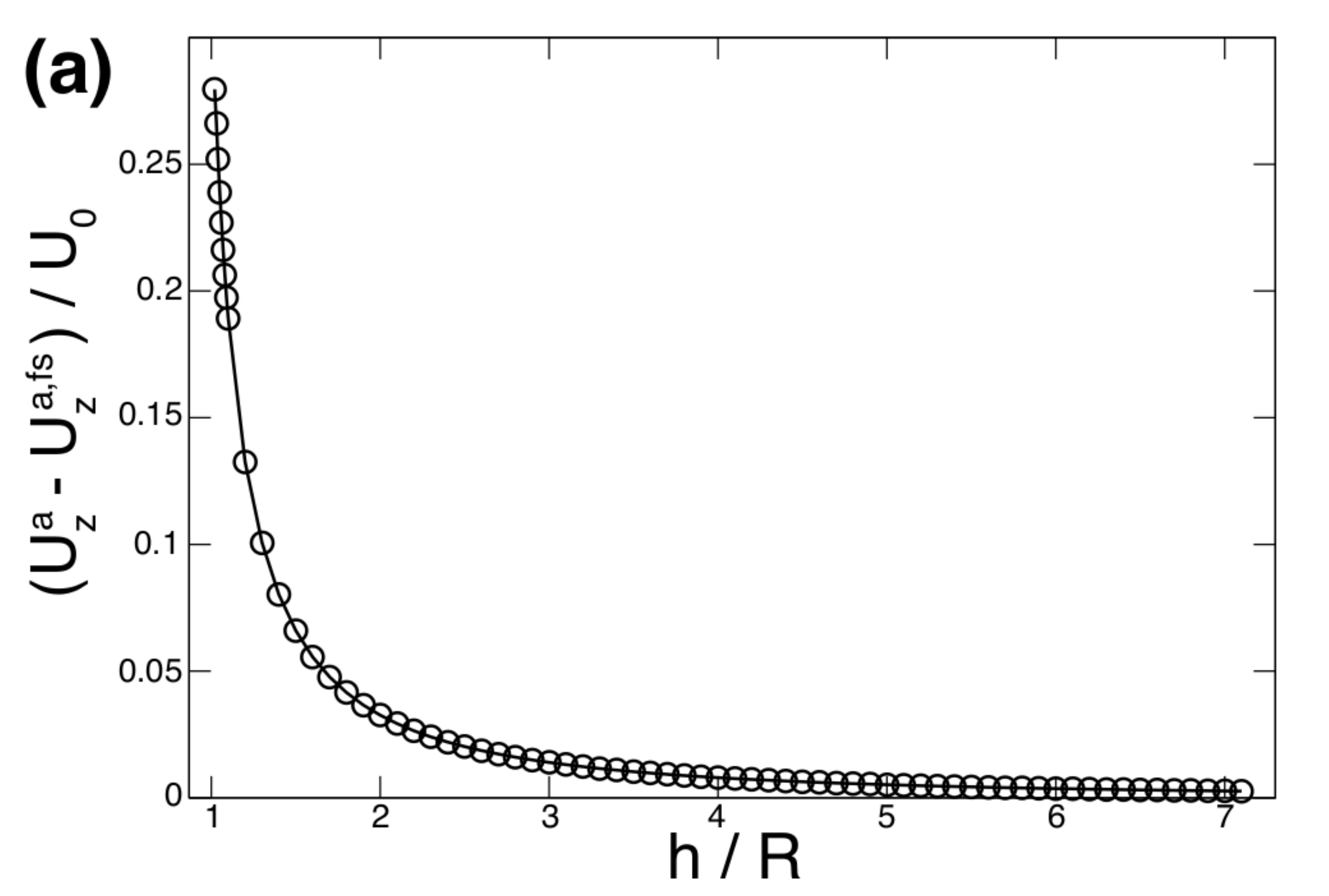}%
\hspace*{0.05\columnwidth}%
\includegraphics[width = 0.45\columnwidth]{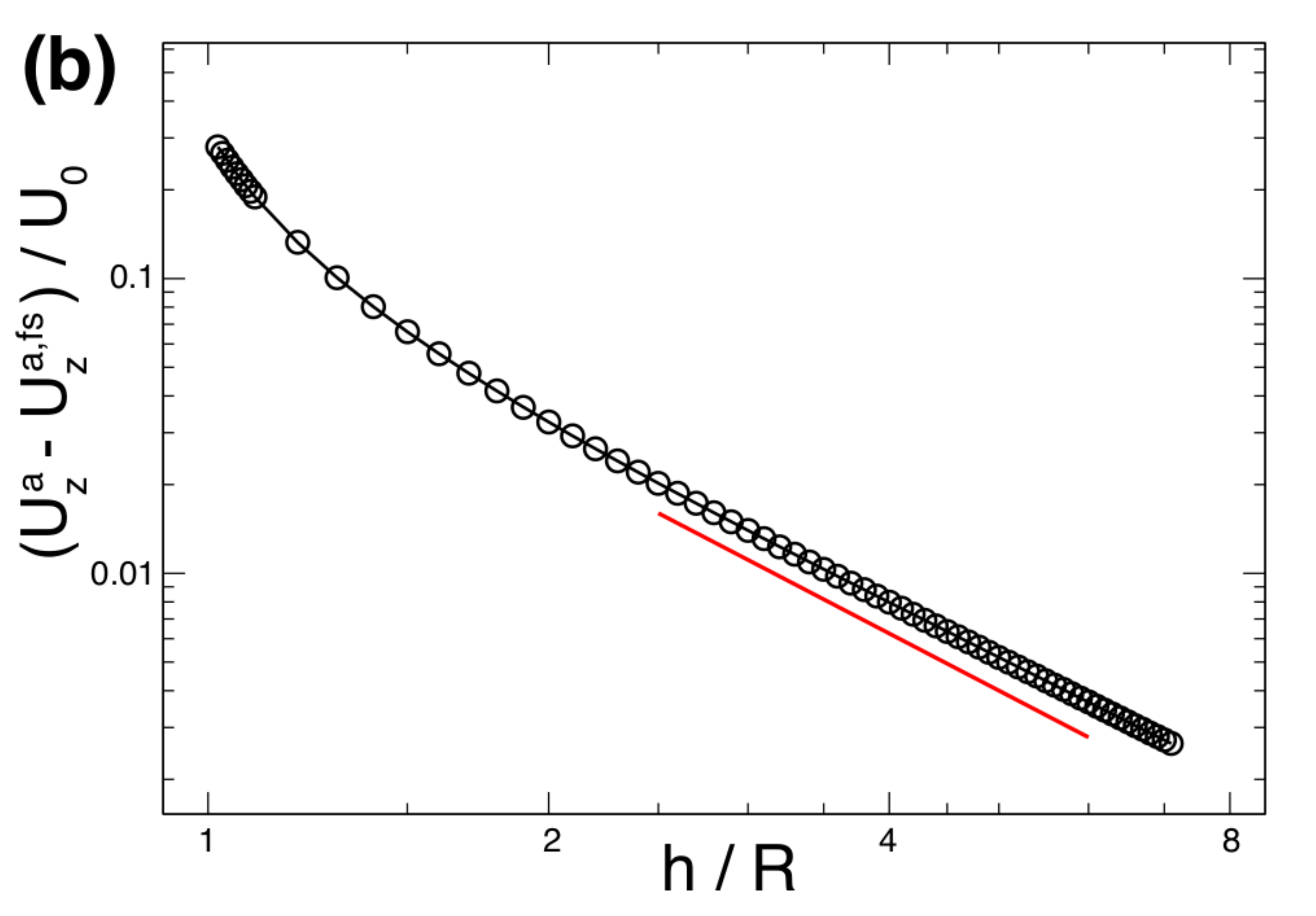}
\caption{\label{SI_fig4} 
Wall-induced change of the particle velocity (a), for a 
particle with its cap oriented towards the wall ($\theta = 0^\circ$). Chemi-osmotic and 
gravitational effects are not included. (b), Same plot as in (a), but with a log-log 
scale. The red line shows a $(h/R)^{-1}$ scaling.
}
\end{figure}

\newpage

\vspace*{1.5in}

\begin{figure}[!htb]
\includegraphics[width = 0.7\columnwidth]{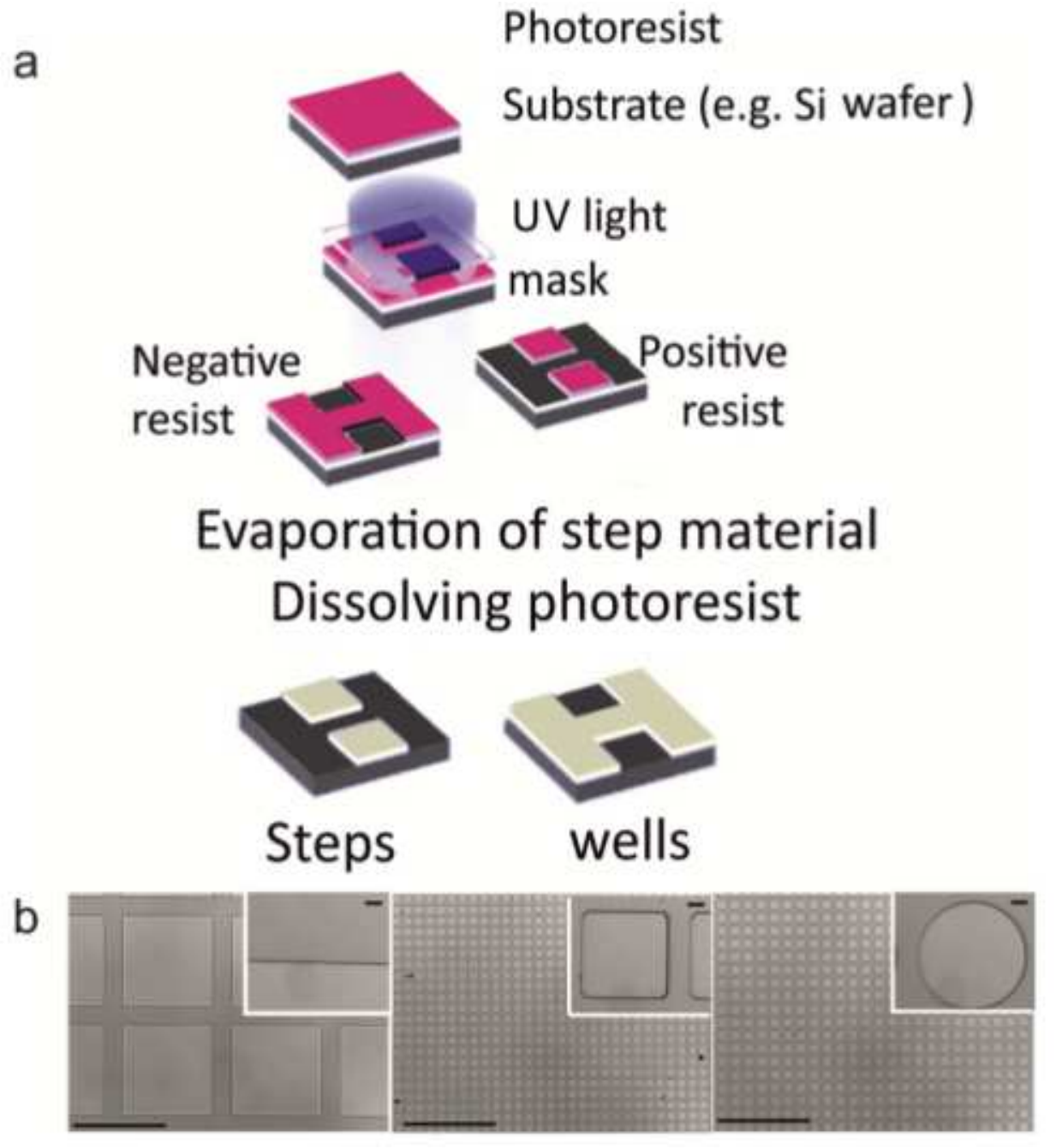}
\caption{\label{SI_fig5} 
Lithography based creation of patterns. (a) Schematic 
representing the lithography based method to create submicron step and well structures 
for particle guidance. (b) Optical images of the resulting patterns; the scale bars in the 
insets correspond to 10 $\mathrm{\mu m}$, while the scale bars in the main images 
correspond to 500 $\mathrm{\mu m}$.
}
\end{figure}

\newpage

\subsection*{{\large Supplementary Tables}}

\vspace*{0.5 in}

\begin{table}[!hpbt]
\begin{center}
  \begin{tabular}{ c | c | c | c | c }
    \hline
    $B_{2}/B_{1}$ & $h_{eq}/R$, no gravity &$\theta_{eq}$, no gravity & $h_{eq}/R$, 
    with gravity & $\theta_{eq}$, with gravity  \\ \hline \hline
    -7 & unstable & unstable & 1.16 and 1.04 (backwards) & 51.5$^{\circ}$ and 
138$^{\circ}$ (backwards) \\ 
    \hline
    -5 & unstable & unstable & 1.19 and 1.03 (backwards) & 49.4$^{\circ}$ and 
142$^{\circ}$ (backwards) \\ 
    \hline
    -3 & unstable & unstable & 1.26 & 42.2$^{\circ}$ \\ \hline
    -1 & none & none & 1.39 & 0$^{\circ}$ \\ \hline
     0 & none & none & 1.24 & 0$^{\circ}$ \\ \hline
     1 & none & none & Below 1.02 & 0$^{\circ}$ \\ \hline
     3 & 1.64 & 102$^{\circ}$ & Below 1.02 & Around 45$^{\circ}$\\ \hline
     5  & 1.22 & 114$^{\circ}$ & Below 1.02 & Around 52$^{\circ}$ \\ \hline
     7 & 1.17 & 117$^{\circ}$ & 1.032 (backwards) & 53.5$^{\circ}$ (backwards) \\ \hline
    \hline
  \end{tabular}
\caption{\label{squirmer_table} 
Attractor states for a squirmer with $R = 2.5 \; 
\mathrm{\mu m}$ and $B_{l} = 0$ for $l > 2$. ``Unstable'' indicates that there is an 
unstable fixed point. ``None'' indicates that there is no fixed point. ``Backwards'' 
indicates that the squirmer (which in the bulk swims ``away'' from the cap) swims 
\textit{towards} its cap when in the steady state near the wall, which is not observed 
experimentally. In some cases (indicated in the table as ``Below 1.02''), there are clear 
signs of an attractor with $h_{eq}/R$ below the numerical cut-off of $h/R = 1.02$; 
however, the corresponding orientations are significantly different from the values 
$\approx 90^{\circ}$ observed in experiments.
}
\end{center}
\end{table}


\begin{table}
\begin{center}
  \begin{tabular}{ c | c | c | c | c || c | c | c | c }
    \hline
   $\beta$ & $B_{1}$ & $B_{2}$ & $B_{3}$ & $B_{2}/B_{1}$ & $h_{eq}/R$, no gravity 
&$\theta_{eq}$, 
   no gravity & $h_{eq}/R$, with gravity & $\theta_{eq}$, with gravity \\  \hline \hline
	-0.9 & 0.019 & 0.413 & 0.033 & 22.04 & 1.05, 1.03 (bwds) & 68.9$^{\circ}$, 
109$^{\circ}$ (bwds) 
	& 1.056, 1.03 (bwds) & 66.8$^{\circ}$, 111$^{\circ}$ (bwds) \\ \hline
    -0.8 & 0.038 & 0.392 & 0.066 & 10.44 & 1.063 & 69.7$^{\circ}$  & 1.09 & 65.3$^{\circ}$ 
\\ \hline
    -0.7 & 0.056 & 0.370 & 0.098 & 6.573 & 1.075 & 70.2$^{\circ}$  & 1.16 & 59.9$^{\circ}$ 
 \\ \hline
    -0.6 & 0.075 & 0.348 & 0.131 & 4.64 & 1.094 & 71$^{\circ}$  & 1.3 & 48.5$^{\circ}$ \\ 
\hline
    -0.5 & 0.094 & 0.326 & 0.164 & 3.48 & $\sim1.15$ (limit cycle) & $\sim72^{\circ}$ 
(limit cycle) 
    & 1.56 & 0$^{\circ}$ \\ \hline
    -0.4 & 0.113 & 0.305 & 0.197 & 2.707 & unstable & unstable & 1.48 & 0$^{\circ}$ \\ 
\hline
    -0.3 & 0.131 & 0.283 & 0.230 & 2.154 & unstable & unstable & 1.42 & 0$^{\circ}$ \\ 
\hline
    -0.1 & 0.169 & 0.239 & 0.295 & 1.418 & none & none & 1.32 & 0$^{\circ}$ \\ \hline
     0 & 0.188 & 0.218 & 0.328 & 1.16  & none & none & 1.29 & 0$^{\circ}$ \\ \hline
     0.1 & 0.206 & 0.196 & 0.361 & 0.949 & none & none & 1.26 & 0$^{\circ}$\\ \hline
     0.3 & 0.244 & 0.152 & 0.427 & 0.625 & none & none & 1.21 & 0$^{\circ}$\\ \hline
     0.7 & 0.319 & 0.065 &  0.558 & 0.205 &  none & none & 1.14 & 0$^{\circ}$ \\ \hline
     1.3 & 0.431 & -0.065 & 0.755 & -0.151 & none & none & 1.09 & 0$^{\circ}$ \\ \hline
     2.0 & 0.563 & -0.218 & 0.984 & -0.387 & none & none & Below 1.02 & 0$^{\circ}$ \\ 
\hline
  \end{tabular}
\caption{\label{effective_squirmer_table} 
Hydrodynamic properties and attractor locations for the effective squirmer with $R = 2.5 \; 
\mathrm{\mu m}$. ``Bwds'' indicates steady motion \textit{towards} the cap, which is not 
observed experimentally. ``Limit cycle'' indicates one case with a sustained oscillation 
of small amplitude in $h/R$ and $\theta$. ``Unstable'' indicates  the presence of an 
unstable fixed point. In one case, there are clear signs of a ``hovering'' attractor with 
$h_{eq}/R$ below the numerical cut-off. Although we listed the amplitudes of the first 
three squirming modes, we note that the effective squirmer can also have $B_{l} \neq 0$ 
for $l > 3$.
}
\end{center}
\end{table}

\newpage

\subsection*{{\large Supplementary Note 1: Calculation of Gravitational Contribution to 
Particle Motion}}

The particle experiences a gravitational force from the weight of the spherical silica 
core and platinum cap, as well as a gravitational torque from the bottom-heaviness 
imparted by the platinum cap. For a given height and orientation of the particle, we use 
the BEM to calculate the hydrodynamic resistance tensor of the particle. We then obtain 
the gravitational contributions $\mathbf{U}^g$ and $\mathbf{\Omega}^g$ to the particle 
velocity as the product of the inverse of this tensor (i.e. the hydrodynamic mobility 
tensor) and the vector containing the six components of gravitational force and torque. 
(Our calculation of the gravitational contribution to velocity therefore includes the 
effect of hydrodynamic interaction with the planar wall.)

The linearity of the Stokes equation allows one to sum the separate contributions of 
activity and gravity to determine the complete translational and angular velocities as 
$\mathbf{U} = \mathbf{U}^a + \mathbf{U}^g$ and $\mathbf{\Omega} = \mathbf{\Omega}^a + 
\mathbf{\Omega}^g$. However, the two sets of velocities must be expressed in the same 
units. The contributions from particle activity are obtained in terms of $U_0 := |b_{cap}| 
\kappa/D$ and $\Omega_0 := U_0/R$, i.e., as $\mathbf{U}^a/U_0$ and  
$\mathbf{\Omega}/\Omega_0$. We estimate $U_0$ and $\Omega_0$ by taking $U_p = 
6~\mathrm{\mu m/s}$ as a typical particle velocity. Within the neutral 
self-diffusiophoresis framework, $U_p/U_0$ can be calculated analytically or numerically 
as a function of the material parameters of the particle, i.e. the extent of catalyst 
coverage and the spatial variation of surface mobility $b(\mathbf{r})$ [SI1,SI2]. For 
instance, $U_p/U_0 = 1/4$ for half coverage and uniform surface mobility [SI3]. For a 
given set of material parameters, we can therefore calculate the characteristic velocity 
$U_0$ in dimensional units from $U_p/U_0$ and $U_p = 6~\mathrm{\mu m/s}$.

\subsection*{{\large Supplementary Note 2: Isolation of contribution to particle motion}}

We seek to isolate and quantify the various physical contributions to the motion of a 
particle. We focus on the $x$-component of the angular velocity $\mathbf{\Omega}$. We 
recall that the linearity of the Stokes equations permits us to solve for 
\textit{a}ctive, \textit{g}ravitational, and \textit{w}all \textit{s}lip (chemi-osmotic) 
contributions separately and superpose them to obtain the full angular velocity: 
$\Omega_{x} = \Omega_{x}^{g} + \Omega_{x}^{a} + \Omega_{x}^{ws}$. To obtain the active 
contribution, we use the Lorentz reciprocal theorem. This theorem allows the problem of 
determination of $(\mathbf{U}^{a}, \mathbf{\Omega}^{a})^{T}$ to be related to six 
``primed'' problems with the same geometry but different boundary conditions. 
We obtain six coupled equations:
\begin{equation}
\label{reciprocalthm}
\mathbf{U}^{a} \cdot \mathbf{F}'_{j} + \mathbf{\Omega}^{a} \cdot 
\boldsymbol{\tau}'_{j} 
= - \int\limits_{particle} 
\mathbf{v}_{s} \cdot \boldsymbol{\sigma}'_{j} \cdot \mathbf{n} \; dS \;,~j = 
1,\dots,6\;,
\end{equation}
where $\mathbf{F}'_{j}$ and $\boldsymbol{\tau}'_{j}$ are the force and torque, 
respectively, 
exerted by quiescent fluid on a particle in steady translation (or rotation) with a 
no-slip boundary condition on its surface. The index $j$ denotes steady translation in 
the $\hat{x}$, $\hat{y}$, or $\hat{z}$ direction for $j = 1, 2, 3$, respectively, or 
steady rotation in $\hat{x}$, $\hat{y}$, or $\hat{z}$ for $j = 4, 5, 6$.  Likewise, 
$\boldsymbol{\sigma}'_{j}$ is the fluid stress tensor for the steadily translating or 
rotating 
particle. Further details concerning the derivation of Eq. (\ref{reciprocalthm}) are 
provided in our previous work [SI4]. We recall that $\mathbf{v}_{s} = -b(\mathbf{r}) 
\nabla_{||} c(\mathbf{r})$. From Eq. (\ref{reciprocalthm}), we find the component 
\begin{equation}
\label{eq_Omega_x}
\Omega_{x}^{a} = \sum_{j=1}^6  (\mathbf{R}^{-1})_{4,j} \int\limits_{particle} 
b(\mathbf{r}) 
\nabla_{||} c(\mathbf{r}) \cdot \boldsymbol{\sigma}'_{j} \cdot \mathbf{n} \; dS.
\end{equation}
Here, the forces and torques  $\mathbf{F}'_{j}$ and $\boldsymbol{\tau}'_{j}$  have 
been compactly organized into a matrix $\mathbf{R}$, which has the $k$-th row 
$\mathbf{R}_k$ 
given by 
\begin{equation}
\label{Rmatrix}
\mathbf{R}_k := \left(\mathbf{F}'_{kx}, \mathbf{F}'_{ky}, \mathbf{F}'_{kz}, 
\boldsymbol{\tau}'_{kx}, \boldsymbol{\tau}'_{ky}, \boldsymbol{\tau}'_{kz} \right)\,.
\end{equation}
The contribution in Eq. (\ref{eq_Omega_x}) to the angular velocity of the 
particle is shown in Fig. \ref{fig3} of the main text. 

We now write the activity-induced solute concentration $c (\mathbf{r}) = 
c^{fs}(\mathbf{r}) + \delta c (\mathbf{r})$ with a {\it f}ree {\it s}pace component 
$c^{fs}(\mathbf{r})$ and the wall correction $\delta c (\mathbf{r})$. Similarly, we write 
$\boldsymbol{\sigma}'_{j} = \boldsymbol{\sigma}'^{fs}_{j} + \delta 
\boldsymbol{\sigma}'_{j}$ and 
$\mathbf{R} = \mathbf{R}_{fs} + \delta \mathbf{R}$. Using these representations in the 
integral in Eq. (\ref{eq_Omega_x}), we may estimate various contributions to the particle 
rotation. \newline

\noindent The free space angular velocity $\Omega_x^{a,fs} := \sum_{j=1}^6  
(\mathbf{R}_{fs}^{-1})_{4,j} \int\limits_{particle} b(\mathbf{r})\nabla_{||} 
c^{fs}(\mathbf{r}) \cdot \boldsymbol{\sigma}_{j}'^{fs} 
\cdot \mathbf{n} \; dS$ is zero, due to the axial symmetry of the particle. \newline

\noindent $\Omega_x^{a,hi} := \sum_{j=1}^6 (\mathbf{R}^{-1})_{4,j} 
\int\limits_{particle} b(\mathbf{r}) \nabla_{||} c^{fs}(\mathbf{r}) \cdot 
\delta \boldsymbol{\sigma}'_{j} \cdot \mathbf{n} \; dS $ gives the 
contribution to $\Omega_x^{a}$ strictly from {\it h}ydrodynamic {\it i}nteractions with 
the wall. It is plotted in Fig. \ref{fig3}(b) of the main text. \newline

\noindent $\Omega_x^{a,sol} :=  \sum_{j=1}^6 (\mathbf{R}^{-1}_{fs})_{4,j}  
\int\limits_{particle} b(\mathbf{r}) \nabla_{||} \delta c (\mathbf{r}) \cdot 
\boldsymbol{\sigma}'^{fs}_{j} \cdot \mathbf{n} \; dS$ gives the contribution to 
$\Omega_x^{a}$, shown in Fig. \ref{fig3}(c) of the main text, strictly from wall-induced 
{\it sol}ute modifications. In other words, $\Omega_x^{a,sol}$ represents 
\textit{phoretic rotation} of the particle from wall-induced concentration gradients. Note 
that this term is non-zero only when $b_{cap} \neq b_{inert}$; in the case considered 
here, $b_{inert}/b_{cap} = 0.3$.\newline

\noindent Finally, the term $\Omega_x^{a, \delta\delta} := \sum_{j=1}^6 
(\mathbf{R}^{-1})_{4,j} \int\limits_{particle} b(\mathbf{r}) \nabla_{||} \delta c 
(\mathbf{r}) \cdot \delta\boldsymbol{\sigma}'_{j} \cdot \mathbf{n} \; dS $ is due to 
higher order coupling between the chemical and hydrodynamic effects of the wall. It is 
depicted in Fig. \ref{fig3}(d) of the main text. Interestingly, it is not necessarily 
small when the particle is close to the wall. \newline

\noindent Now we turn to the other contributions to $\Omega_{x}$. We show the 
contribution from gravitational torque in Fig.~\ref{fig3}(e) of the main text: 
$\Omega_x^{g}\equiv \mathbf{R}^{-1} \mathbf{\frak{F}}^{g}$, where we define a generalized 
gravitational force $\mathbf{\frak{F}}^{g} \equiv (0, 0, F_{z}^{g}, \tau_{x}^{g}, 0, 
0)^{T}$. This term depends on the size $R$ of the particle; here, it is calculated for $R 
= 2.5\;{\mu m}$.\newline

\noindent Likewise, in Fig.~\ref{fig3}(f) of the main text, we show the contribution from 
\textit{w}all \textit{s}lip, i.e., activity-induced chemio-osmotic flow 
along the wall: $\Omega_x^{ws}\equiv  \sum_{j=1}^6 (\mathbf{R}^{-1})_{4,j}  
\int\limits_{wall} b_{w}(\mathbf{r}) 
\nabla_{||} c \cdot \boldsymbol{\sigma}'_{j}  \cdot \mathbf{n} \; dS $. This component is 
absent in the squirmer model, but it is significant in our case, where 
$b_{w}/b_{cap} = -0.2$.

\subsection*{{\large Supplementary Note 3: Hydrodynamics-only models}}

We turn to the wider question of whether the ``squirmer'' model, in which the interaction 
with the wall is purely hydrodynamic, can reproduce the experimentally observation for 
\textit{some set of parameters}. We recall that the slip velocity of an axisymmetric 
squirmer can be written as:
\begin{equation}
\mathbf{v}_{s}(\theta_{p}) = \sum_{l = 1}^{\infty} B_{l} V_{l}(cos(\theta_{p})) 
\hat{\theta}_{p},
\end{equation}
where $\theta_{p}$ is an angle defined with respect to the axis of symmetry, $V_{l}(x) = 
\frac{2 \sqrt{1 -x^{2}}}{n(n+1)} \frac{d}{dx} P_{l}(x)$, $P_{l}$ is the Legendre 
polynomial of order $l$, and $B_{l}$ is the amplitude of squirming mode $l$ [SI5].

An exhaustive search through the squirming mode amplitudes $B_{l}$ is beyond the scope of 
this work. Nevertheless, we can gain some insight by restricting our consideration to the 
first two squirming modes. The amplitude of the first mode is set by the free space 
swimming velocity $v_{f.s.} \approx 6 \; \mathrm{\mu m/s}$, since $v_{f.s.} = 
(2/3)\;B_{1}$. We vary the ratio $B_{2}/B_{1}$, taking $B_{l} = 0$ for $l > 2$, and 
determine whether a sliding state emerges (i) \textit{in the absence of gravity} and (ii) 
\textit{in the presence of gravity}, which respectively represent (i) swimming near a side 
wall, and (ii) swimming above a substrate. The results are shown in Table 
\ref{squirmer_table} for $R = 2.5\,\mathrm{\mu m}$. For situation (i), our results show 
good agreement with Gaffney and Ishimoto [SI5], including the finding that sliding states 
emerge only for $B_{2}/B_{1} 
\ge 3$.

We find that, for the parameters considered, the squirmer model cannot reproduce the 
experimental observation that a particle swims at $\theta \approx 90^{\circ}$ in 
\textit{both} situation (i) and situation (ii). Gravity shifts the sliding states that 
occur for $B_{2}/B_{1} \geq 3$ to $\theta_{eq} \approx 45^{\circ}$. This steady angle 
would be detectable experimentally as a large apparent coverage of the particles by 
catalyst.  We conclude that, for the parameters considered in Table \ref{squirmer_table}, 
the effect of the force dipole (the strength of which is proportional to the amplitude 
$B_{2}$ of the second squirming mode) is \textit{too weak} to balance the effect of 
gravity at $\theta \approx 90^{\circ}$.

In constructing Table \ref{squirmer_table}, we made two simplifying and physically 
plausible assumptions: 1.) The contributions of the higher order modes ($l > 2$) to the 
disturbance velocity decay rapidly with distance from the swimmer, and hence contribute 
negligibly to interaction with the wall; and 2.) the ratio $B_{2}/B_{1} \sim 
\mathcal{O}(1)$. We can relax both assumptions by considering an ``effective squirmer'' 
obtained within our model for a self-diffusiophoretic swimmer. The effective squirmer is 
obtained for a given $\beta = b_{inert}/b_{cap}$ by neglecting the effect of the wall on 
the concentration field of a self-diffusiophoretic particle, i.e., by using $c_{f.s.}$ as 
described in Supplementary Note 2. Additionally, chemi-osmotic effects are 
neglected. An effective squirmer could have, in principle, non-zero amplitude for all 
$B_{l}$. Secondly, as shown in Table \ref{effective_squirmer_table}, the ratio 
$B_{2}/B_{1} \rightarrow \infty$ as $\beta \rightarrow -1$ from above. This is because, 
for $\beta \approx -1$, one face of the particle is attracted by solute, and the other 
repelled by solute; furthermore, the strengths of attraction and repulsion are 
approximately equal. $B_{1}$ is therefore nearly zero, since it is proportional to the 
velocity of the particle.

Our results are in Table \ref{effective_squirmer_table}. We find that, for the parameters 
studied, the effective squirmer cannot reproduce the experimental observations in both 
situation (i) and situation (ii). The closest match is for $\beta = -0.8$, with 
$\theta_{eq} \approx 70^{\circ}$ in situation (i) and $\theta_{eq} \approx 65^{\circ}$ in 
situation (ii). Achieving this particular sliding state requires a very strong force 
dipole interaction with the wall: $B_{2}/B_{1} \approx 10$. Is this physically plausible? 
A recent estimate of the force dipole strength for a catalytic Janus particle 
was provided by Brown \textit{et al.} [SI6]. For a $R \approx 1\; \mathrm{\mu m}$ colloid 
that moves at $U \approx 15 \; \mathrm{\mu m/s}$, they estimate $\alpha = 30 \; 
\mathrm{\mu m^{3}/s}$. Non-dimensionalizing, and then using the expression from Gaffney 
and Ishimoto that connects $B_{2}/B_{1}$ with dimensionless $\alpha$ [SI5], we find 
$B_{2}/B_{1} \approx 2.5$.  Hence, the $\beta = -0.8$ effective squirmer is both 
physically unlikely (having a very large $B_{2}/B_{1}$, i.e., a very large force dipole) 
\textit{and} a poorer fit to experimental observations than the full model presented in 
our work.  For more realistic values of the force dipole, hydrodynamics is \textit{too 
weak} to, by itself, balance gravitational effects at $\theta \approx 90^{\circ}$.

\subsection*{{\large Supplementary Note 4: Range of interaction with the wall}}

The interaction of a chemically active particle with a planar wall has a long-ranged 
character. However, we show that the amplitude (strength) of this interaction is very 
small except when the particle is close to the wall. This makes it difficult to detect 
its effects experimentally (at least with the equipment and techniques currently 
available to us).

In Fig. \ref{SI_fig4}(a), we plot the wall-induced change of the wall normal component of 
the self-diffusiophoretic velocity of a particle that has its cap oriented towards 
the wall ($\theta = 0^{\circ}$). The subtraction of the free space self-diffusiophoretic 
velocity $U_{z}^{a,fs} = 0.1625\,U_{0}$ from $U_{z}^{a}$ isolates the effect of the wall. 
(Note that chemi-osmotic and gravitational contributions are not included in Fig. 
\ref{SI_fig4}(a).)  At $h/R = 5$, the contribution to $U_{z}^{a}$ from the wall has 
already decayed to approximately three percent of the free space self-diffusiophoretic 
velocity. This change in speed is too small to be apparent when viewing an optical 
microscopy video of a particle near a step or a side wall. The effect of the wall is even 
weaker for orientations $\theta > 0^{\circ}$.

Our numerical calculations recover the long-ranged character of the interaction. In 
Fig. \ref{SI_fig4}(b), we show the same plot as in (a), but with a log-log 
scale. Far away from the wall, $\Delta U_{z}^{a}$ follows a $(h/R)^{-2}$ power law, which 
is shown as a red line. The effect of the wall on the solute field decays as $1/r$, i.e., 
the leading order term representing the wall is an image point source. Since $\Delta 
U_{z}^{a}$ is proportional to the wall-induced concentration gradient, it decays as 
$1/r^{2}$.

\subsection*{{\large Supplementary Note 5: Robustness of the sliding state}}

Finally, we comment on the robustness of the sliding states against thermal fluctuations, 
which were not included in our model. To this end we perform a standard linear stability 
analysis of our dynamical system, which may be written as $\{\dot h = f_1(h,\theta), \dot 
\theta = f_2(h,\theta)\}$, at a fixed point $(h_{eq},\theta_{eq})$ at which 
$\{f_1(h_{eq},\theta_{eq}) = 0, f_2(h_{eq},\theta_{eq})=0\}$. This amounts to 
determination of the eigenvalues $\lambda_{1,2}$ of the Jacobian matrix $\partial 
[f_1,f_2]/\partial [h,\theta]$ evaluated at $(h_{eq},\theta_{eq})$.

For the fixed point at Fig. \ref{fig2}d of the main text, we obtain that the Jacobian has 
eigenvalues $\lambda_{1,2} \simeq (−0.40 \pm 0.17\, i)~U_0/R$. Since the real part of 
the eigenvalues is negative, the fixed point is a stable attractor: a small perturbation 
away from the fixed point will exponentially decay with a characteristic timescale 
$\tau = 1/\mathrm{Re}(\lambda_1)$. To convert this timescale into dimensional units, we 
use $R = 2.5\; \mathrm{\mu m}$ and $U_0 \approx 6.2~U_{f.s.} = 6.2 \times 6~ \mathrm{\mu 
m/s}$, where $U_{f.s.}$ is the self-propulsion velocity of a half-covered Janus swimmer 
with $b_{inert}/b_{cap} = 0.3$ in free space [SI2]. We obtain the timescale $\tau 
\approx 0.37$ s for the self-trapping of a particle into this sliding state. For 
comparison, the characteristic timescales of rotational and translational diffusion are 
$\tau_r \approx 97$ s and $\tau_t \approx 70$ s, respectively. This separation of 
timescales indicates that the sliding state in Fig. \ref{fig2}d of the main text is 
robust against thermal noise. Similarly, for the fixed points in Fig. \ref{fig2}f of the 
main text and \ref{SI_fig2}, we obtain characteristic self-trapping timescales $\tau 
\approx 1.7$ s $\tau \approx 0.4$ s, respectively (in the latter case we use $R = 1.0\; 
\mathrm{\mu m}$ for the dimensionalization).

\vspace*{0.5 in}

\noindent \textbf{{\large SI References}}\newline\newline
[SI1] Golestanian, R., Phys. Rev. Lett. \textbf{102}, 188305 (2009).\newline
[SI2] Golestanian, R., Liverpool, T. B., and Ajdari, A.,  New Journal of Physics 
\textbf{9}, 126 (2007).\newline
[SI3] Popescu, M. N., Dietrich, S., and Oshanin, G., J. Chem. Phys. \textbf{130}, 194702 
(2009).\newline
[SI4] Uspal, W. E., Popescu, M. N., Dietrich, S., and Tasinkevych, M., Soft Matter 
\textbf{11}, 434 (2015). \newline
[SI5] Ishimoto, K., and Gaffney, E. A., Phys.Rev. E \textbf{88}, 062702 (2013).\newline
[SI6] Brown, A. T. \textit{et al.}, Soft Matter \textbf{12}, 131 (2016).\newline

\end{document}